\documentclass[preprint]{aastex}

\newcommand{\kmsM} {km~s$^{-1}$~Mpc$^{-1}$}
\newcommand{\subsun}{\mbox{$_{\odot}$}}
\newcommand{\etal}{{\it et al.\/}}
\begin{document}

\title{Star Formation and X-ray Emission in Distant Star Forming 
Galaxies\altaffilmark{1}}

\author{Judith G. Cohen\altaffilmark{2}}

\altaffiltext{1}{Based in part on observations obtained at the
W.M. Keck Observatory, which is operated jointly by the California 
Institute of Technology, the University of California, and the
National Aeronautics and Space Administration.}

\altaffiltext{2}{Palomar Observatory, Mail Stop 105-24,
California Institute of Technology, Pasadena, Ca., 91125, jlc@astro.caltech.edu}

\begin{abstract}
About 45\% of the point sources detected in the 2 Ms Chandra exposure
of the HDF-N can be matched with moderately bright galaxies with
$z<1.4$ that have been studied by the Caltech Faint Galaxy Redshift 
Survey.  Although the optical spectra of these galaxies appear normal,
based on their X-ray properties $\sim$20\% of them appear to
contain weak AGNs. More than 90\% of the X-ray photons detected
by Chandra from galaxies within the  redshift regime $0.4 < z < 1.1$
are powered by accretion onto massive
black holes. For the sample of galaxies in common,
we use their emitted luminosity in the 3727~\AA\
line of [OII] to estimate their star formation rate (SFR).
The X-ray emitting galaxies are not those with the highest
rest frame equivalent width in this emission line, but rather
are among those with the highest SFR.
With SFR corrected for inclination effects,
the distant galaxies show a $L_X$ -- SFR relationship that is 
comparable to that of local galaxies.  The HDF sample has
a significantly higher median SFR and median SFR/galaxy stellar
mass than does a sample of local
star forming galaxies.
We demonstrate that the observed SFR for most of the galaxies
at $z \sim1$
in the HDF sample, if maintained as constant over their ages, 
suffices to produce the stellar mass observed in these galaxies.
A rise in SFR at still earlier times is not required.
We provide further evidence
to support the conclusion that, once AGNs are eliminated, 
X-ray emission in these distant
star forming galaxies is related to the SFR  through
the same physical mechanisms that prevail locally. 

\end{abstract}


\section{Introduction} 

The release of the point source catalogs for the
2 Ms exposure of the Chandra Deep Field-North by
\cite{alexander03} represents a milestone for
astronomy.  The large collecting area combined with high
spatial resolution of Chandra have, with this extremely long
exposure time, achieved sufficient sensitivity
to detect the population of distant normal galaxies.
Much previous work on the optical counterparts to the
Chandra detections in the region of the HDF,
such as that of \cite{barger02} and \cite{barger03}, has focused
on the optically faintest counterparts of X-ray sources
in efforts to find very distant galaxies, heavily obscured
distant AGNs, etc.
In this paper we address the origin of the X-ray emission
from the Chandra sources with  optical
counterparts corresponding to ``faint'' field galaxies using material from the
Caltech Faint Galaxy Redshift Survey (CFGRS),
described in \cite{cohen00} and \cite{cohen01}.

The CFGRS was 
carried out in
the region of the Hubble Deep Field-North (HDF)
\citep{williams96}, which is included within
the larger area of the CDF-N.
Our redshift survey there achieves very high
completeness to a deep limiting magnitude. The survey sample
was selected based on the four-color
optical and infrared photometric catalogs 
of \cite{hogg00} and covers the area within a diameter
of 8 arcmin centered on the HDF.
Objects were observed spectroscopically using LRIS at the
10-m Keck Telescope \citep{oke95}
irrespective of morphology to include all AGNs, QSOs
and other possible extragalactic objects that might appear stellar.
Our composite catalog for the region of the
HDF contains $\approx$595 galaxies with $z < 1.5$ and includes the published
data from \cite{phillips97}  (from the Lick Deep Group) and the
work of the Caltech and of the Hawaii group published in 
\cite{cohen96} and subsequently to 1999,
as well as five unpublished redshifts from C. Steidel.
Our composite catalog
contains redshifts for 
more than 93\% of the sample to a limiting magnitude of
$R < 24$ in the HDF itself, and to $R < 23.5$ in the Flanking
Fields.

Of the 503 Chandra sources securely detected by 
\cite{alexander03}, 149 are within the 8 arcmin
diameter area covered by the CFGRS.  We adopt a requirement that
the X-ray and optical positions must match to within a tolerance
of 1.5 arcsec
based on the astrometric accuracy of the Chandra and 
the \cite{hogg00} catalogs. We then find that
67 of these X-ray point sources can be matched with
galaxies with $z < 1.4$ in this 
field.\footnote{One additional Chandra point
source is matched with a
Galactic M dwarf star, and one other is matched with
a high redshift QSO in this field.}
Hence about 45\%
of the very deep X-ray detections reported by \cite{alexander03}
arise from galaxies at intermediate redshifts
which appear from their optical
spectra to be  normal.  In the present paper
we compare the X-ray emission detected by Chandra with
the star formation rates inferred from the [OII] emission lines
of these galaxies.   As in earlier papers in this series, we 
adopt the cosmology H$_0 = 60$ \kmsM, $\Omega_M = 0.3$, 
${\Omega}_{\Lambda} = 0$.
Over the redshift interval of most interest, a flat universe with
$\Omega_{\Lambda} = 0.7$ and a Hubble constant of H$_0 = 67$ \kmsM\
gives galaxy luminosities very close to those used below.

\section{Eliminating AGNs from the Sample}

The mechanisms giving rise to the
X-ray emission of local star forming galaxies,
reviewed by \cite{fabbiano89}, 
are a composite of emission from the ISM, from
stellar sources (predominantly X-ray binaries),
from supernova remnants, and possibly from a nuclear
AGN.  With the aid of Chandra's high spatial
resolution,  separation  of these components
in nearby galaxies becomes feasible, see
e.g. \cite{zezas02} for the Antennae galaxy.
In nearby galaxies with high current SFR,
X-ray binaries provide about half of the
flux in the full Chandra band.  The hot components of the ISM,
which contribute most of the remaining flux, can also be studied
in detail with Chandra. The superbubbles, whose evolution was followed
by \cite{tenorio99}, can be resolved, and galactic winds
can be detected and the metallicity of the gas within
them determined, e.g. \cite{martin02}.
In addition,  the sensitivity
of XMM permits detection of a complex array of lines
and precision spectral fitting for X-ray bright nearby
galaxies such as M82 \citep{read02}.
While \cite{horn03} have detected off-nuclear luminous X-ray
sources in galaxies to $z \sim 0.1$ in the HDF, 
none of this richness can be achieved at present for the distant galaxies
in our sample.

The variation of the X-ray emission from nuclear sources,
particularly given the low spatial resolution of early X-ray images,
can be explained by either
the XRB or the AGN hypothesis. \citep*[See, for example,][for M982.]{ptak99}
For M51, however,
the detection of optical emission lines characteristic
of a central weak low luminosity AGN 
by \cite{ho97} is definitive proof that weak AGNs may
be present even in these heavily star forming galaxies; they
find  such weak AGN in a high fraction of local galaxies.
But the nucleus of M51 contributes only 12\% of its total
full band Chandra luminosity; the bulk of its X-ray emission
comes from the extended and stellar sources \citep{terashima03}.

We consider here only those galaxies in the region of the HDF
whose optical spectra show signs of moderate to strong current 
star formation, i.e. galaxies with easily detected optical emission
lines.  These have been assigned
galaxy spectral classes in the CFGRS catalog
using the definition of \cite{cohen99}
of $\cal{I}$ (moderately strong emission lines with detectable
absorption lines as well) 
and $\cal{E}$ (very strong emission lines).  Two  
broad lined AGNs have been culled out of the sample of 595 
galaxies; both of these have very high X-ray luminosities
from the Chandra data.  However, elimination of
narrow-lined AGN 
based solely on their optical spectra is difficult as, for the
higher redshift galaxies considered here, none of the key 
diagnostic features
fall within the wavelength coverage of the available optical spectra.
Furthermore, the light of any nuclear AGN will be diluted in such distant
galaxies by the large metric aperture used for 
ground based spectroscopic observations
\citep{moran02}.  

In the local Universe, the mass of a nuclear
AGN (i.e. a black hole) depends primarily on the mass of the 
stellar bulge component of the host galaxy \citep{ferrarese00}.
The luminosity of the star forming galaxies in the
HDF is, for a given star formation rate (SFR),
in the mean considerably higher than that for
local star forming galaxies \citep{cowie96,cohen01}.
Furthermore, at high redshift, there may be more gas to
power the central source.  Hence
we may expect the impact of contributions from
AGNs to be more important in the HDF than in local samples.

There is a well established relationship between the
far-IR luminosity and the radio luminosity
\citep*[see, e.g.][]{condon92} among 
local star forming galaxies, whose origin
was discussed by \cite{bell03}.  
\cite{garrett02} explored this relation at moderate
redshifts by comparing the far-IR (from ISO)
and radio emission of 20 galaxies from the
CFGRS in the HDF with a local sample.
He found
that the correlation between far-IR and radio
emission established in local star forming galaxies
continues to apply out to $z \sim 1.3$.  Deviations
of excess radio
luminosity  
from such a well defined relationship  presumably
imply the presence of an AGN. However, the
far-IR data in the region of the HDF \citep*[from ISO,][]{aussel99}
are quite limited both in depth and in size of field
and are insufficient for our purposes.
Until SIRTF data are in hand,
we cannot use this approach to test for AGNs.

To shed light on the possible presence of AGNs in our HDF samples,
we build upon the work of \cite{grimm03}, who studied a small sample
of galaxies in the HDF with both radio and
Chandra observations and with redshifts from the CFGRS.
They too suggested that the local
relations  apply to their HDF sample. 
We therefore construct the set of objects which are included in each of
the Chandra point source catalog, the CFGRS (for redshifts),
and the VLA database for the HDF of \cite{richards98}
and \cite{richards00}.  
As one step towards eliminating AGNs, we omit all sources
where the Chandra spectral index is negative (i.e. very hard sources).
Galaxies which are in close pairs of comparable luminosity
in the CFGRS were included 
only if it was clear  to which
component the Chandra detection should be assigned.  If not,
such galaxies were ignored.  Eleven galaxies meet these
criteria.  These galaxies are indicated in Table~\ref{table_sfr}, 
and some median
properties of the sample are given in Table~\ref{table_median}.

Both the X-ray and the radio emission arising from star forming galaxies
are essentially free of reddening effects, and hence
should be tightly coupled. 
Figure~1 shows the resulting $L$(1.4Ghz) versus $L_X$ for
the 11 sources in common.  A $k$-correction for the radio has been
used since all but one of the galaxies have measured radio spectral indices,
but none was used in calculating the  X-ray luminosity.
Power law SEDs, in the form of $f(\nu) \propto \nu^{-\alpha}$,
have $k$-corrections for luminosities ${\nu}L(\nu)$ of 
$(1+z)^{\alpha-1}$.  For a slope $\alpha$ between 1 and 2, 
typical of these X-ray sources, the 
$k$-correction
has a total range from 0.0 to 0.3 dex for $z=0$ to $z=1$.

Most of the points define a tight line, as expected, but two
of them, F36422\_1545\footnote{Galaxy names are based on
the J2000 coordinates of the object; 12 12 34.5 +62 56 43 is identified as
F12345\_5643.}
and F36517\_1220, lie significantly above it.
Luminous discrete non-nuclear X-ray sources, also known as 
ULX sources, found in local galaxies
are reviewed by \cite{makishima00}; the most luminous
of these reach $L_X \ge 4 \times10^{32}$ W in M51 \citep{terashima03}.
These two HDF galaxies have luminosities far higher than any local ULX,
and presumably contain luminous AGNs.  
The redshift of F36422\_1545 is $z=0.857$, so the region of H$\beta$
is too far to the red to be accessible.  The more modest redshift
of F36517\_1220 ($z=0.401$) permits an examination of the 
rest frame 5000~\AA\
region, although H$\alpha$ is inaccessible.  This means that the
conventional AGN diagnostics used in optical spectroscopy of
\cite{baldwin81} and \cite{veilleux87} are out of reach.
Judging by the more limited test of line strength ratios,
avoiding the H$\alpha$ region, introduced by \cite{rola97}, F36517\_1220
is a normal star forming galaxy.  Reconciliation of these conflicting
claims is possible with the conjecture
of \cite{moran02} regarding the importance of aperture effects
on the detection of AGN
at high redshift from optical spectra.

Figure~\ref{fig_xcounts} shows the full-band Chandra counts
as a function of the galaxy redshift
for the sample of star forming
galaxies in the region of the HDF
of this paper.  A detailed description of the sample selection
will be given in \S\ref{section_hdf}.  
We ascribe the sprinkling of galaxies along the bottom of this
figure to X-ray emission associated with normal star formation.
The two suspected AGNs isolated from Figure~\ref{fig_agn} are circled, and
among the most discrepant galaxies in Figure~\ref{fig_xcounts}.
There are two galaxies  in our sample with full band
Chandra  detections which exceed 1000, while the two broad lined AGN
that were eliminated from the sample have full band Chandra detections
in excess of 2000 counts.  

If we assume that all galaxies with
full band Chandra counts in excess of 100 contain a AGN, then
there are two broad lined and 8 narrow lined AGN in our sample,
of which
only the two broad lined galaxies were picked out from their optical
spectra as AGNs.  The fraction of the total Chandra detected flux
originating from accretion onto a massive black hole is 
$\sim$93\%.  The much more numerous normal star forming galaxies
contribute very little to the total X-ray luminosity density.

There are, in addition,
5 to 6 additional galaxies in this sample of 38 which are clearly
discrepant, with excess X-ray flux, though their optical spectra
appear normal.  Only one of these 6 galaxies has a radio detection.
These too presumably have weak AGN nuclei.

To within
the small number statistics of the present sample, we
therefore assert that about
20\% of the spectroscopically normal high luminosity star forming
galaxies in the HDF have nuclear AGNs which contribute
substantially to their total X-ray fluxes.   
\cite{huchra92} gives
the frequency of Seyfert I and Seyfert II galaxies in the
CfA Redshift Survey of the local Universe, while \cite{ho97} have
done so for  AGNs reaching to much lower luminosities.
Both the AGN luminosity and
the very rough AGN fraction derived above for our sample lie
in between these local determinations.  \cite{kobulnicky03}
find that 15\% of their sample of 66 star-forming 
galaxies with $0.26 < z < 0.82$ in the Groth strip are probably narrow
lined AGNs.

$L(X)$, $L($radio),
and $L($FIR) are to first order independent of internal
reddening within the galaxy, and hence
these correlations, including that shown in Figure~\ref{fig_agn}
for a sample of galaxies in the region of the HDF, 
have quite
small dispersions, both among local galaxies and in the HDF
using the CFGRS database,
ignoring the obvious occasional 
outliers, which we presume to be AGN.  The underlying
variable driving these correlations is widely assumed
to be the SFR.

\section{The Sample of Local Calibrators}

We first create a sample of local star forming/starburst
galaxies for comparison with our sample of much more distant
galaxies in the region of the HDF. Their
properties, and the references from which
their X-ray luminosities and H$\alpha$ fluxes were taken, are given
in Table~\ref{table_local}. We have modified the published data to
correspond $H_0 = 70$ \kmsM\ and have removed the Galactic extinction
using the maps of \cite{schlegel98}
when necessary.  We correct the SFR for inclination
effects; the result is denoted SFR$^i$.    We take 
axis ratios for our sample of local calibrators
from the 2MASS Large Galaxy Atlas
\citep{jarrett03}, or, if necessary, measure them
from NED images. We apply a correction of the form
$\Delta$(mag) = $\gamma {\rm{log}}(b/a)$, and adopt
$\gamma$(H$\alpha$) = 1.3 \citep{tully98}.  This does not 
remove the reddening experienced by a face-on disk galaxy.

We have also applied small corrections to 
total  galaxy X-ray fluxes taken from the catalog of galaxy
observations with the Einstein satellite of
\cite{fabbiano92}, given over the regime 0.2 -- 4.0 keV,
to cover the Chandra full bandpass.  The observed fluxes in
the optical emission lines for these local galaxies have in
most cases been measured through apertures large enough to
encompass the entire galaxy, except for the data taken
from \cite{calzetti95}.  Substantial aperture corrections for 
this data set are required and 
were calculated using images from NED 
under the assumption that the spatial distribution of the
flux in the emission line is the same as that of a nearby
broad band continuum-dominated filter.  The
H$\alpha$ fluxes from \cite{calzetti95} as published are corrected for
the internal reddening of the galaxy, and this was backed
out prior to use.

Our local sample of star forming galaxies was selected
based on our ability to locate the required observational
material in the literature.  Few surveys of large aperture observations
of nearby galaxies in the optical emission lines are
available. We began with the sample of \cite{bell01},
and tried to locate the requisite X-ray emission for
the galaxies in their sample  with substantial SFR. 
A special effort had to be made to find suitable local galaxies with
high SFR; such galaxies are rare in the local Universe.  
The two local galaxies with the highest SFR are
at distances $D$ of
56 and 78 Mpc, while all the others have $D < 20$ Mpc, and 
9 of the 14 have $D \le 10$ Mpc.
\cite{kauffmann03}
have analyzed a large sample of galaxies from the SDSS and find that
the percent of local galaxies with signs of a recent or ongoing starburst
decreases rapidly as the luminosity increases, so our difficulty
in finding luminous local galaxies with high SFR is not surprising.

These local galaxies, with H$\alpha$ used as the
diagnostic for star formation, are indicated by large 
filled circles in
Figure~\ref{fig_local_inclincor}; for $(B-V)_0 > 0.6$, 
they are indicated by stars.
A least squares linear fit to the 14 points
is shown by the thick line, and has the form
log($L_X) = 32.783(\pm0.115) + 1.100(\pm0.140) 
{\rm{log}}$[SFR$^i$(H$\alpha$)], with $\sigma$ about this fit
of 0.31 dex. The scatter would be
noticeably larger if the inclination corrections were not used.
Several smaller sources contribute to the remaining scatter
in the relation between $L_X$ and SFR$^i$(H$\alpha$).
They include the small corrections
applied to intercompare observations
made with three different X-ray telescopes
(Chandra, ASCA and Einstein), ignoring the X-ray k-corrections
and, most importantly, ignoring   
reddening differences from galaxy to galaxy within the local sample
after being corrected to face-on.  Table~\ref{table_disp}
gives an estimate of the contribution of each of these terms
to the observed $\sigma$.
In addition, at the lowest $L_X$
for these nearby galaxies, the number
of detected discrete non-nuclear
X-ray sources, presumably various types
of luminous X-ray binaries, which may contribute an amount equal to the
extended emission,
becomes fairly small, and Poisson statistics
may become important.

We are using the 3727~\AA\ emission line of [OII] as our
diagnostic of SFR for our sample of distant galaxies in the 
region of the HDF.  However, among local galaxies
there are fewer measurements of the observed integrated
flux in the 3727~\AA\ emission line of [OII] than there
are for H$\alpha$.  This is presumably
because the [OII] emission line is less dominant
in its spectral region than is H$\alpha$.  Also the  adjacent continuum
is chopped up by strong absorption features there, hence narrow
band imaging is not feasible. 

The major
surveys for the strength of the the 3727~\AA\ emission line of [OII]
among local galaxies are those of
\cite{gallagher89} and \cite{kennicutt92}.
Substantial aperture corrections are required for the former,
while for the latter, the [OII] observed line fluxes are given
with respect to the observed H$\alpha$ flux.
We were able to
locate suitable measurements for only five of the galaxies
in the local sample, all from \cite{kennicutt92}.
The resulting deduced SFR$^i$ values, 
using the calibration of \cite{kennicutt98},
are shown in Figure~\ref{fig_local_inclincor} as open circles.  Thin
horizontal lines connect the H$\alpha$ and [OII] SFRs.
This offset in SFR$^i$ represents mainly the difference $\Delta(A)$
between $A$(H$\alpha$) and $A$(3727),
where $A(\lambda$) is the total absorption for the integrated
light of
each of the calibrating galaxies (corrected to face-on). 

Reddening curves for galaxies
are quite controversial \citep{calzetti99}, but most of the
disagreement lies within the rest-frame UV, which we do not use.
We therefore adopt 
$A(\lambda)/A(V)$ of  \cite{schlegel98}, so
$\Delta(A) = 2.26E(B-V)$ mag. 
The differential extinction appears small for three of the galaxies,
and much larger for NGC~3034 and NGC~5194.  The dashed line
is the fit to H$\alpha$ translated by $-0.7$ dex in log(SFR$^i$)
so as to fit through the values of these last two galaxies, which
are among the more reddened of the local calibrators.  Note that
$\Delta(A) = 0.7~{\rm{dex}} \equiv 1.75$ mag 
corresponds to a total reddening for a face-on galaxy of
$E(B-V)$ = 0.8 mag.  This is not unreasonable for 
star forming galaxies; 
the references quoted in Table~\ref{table_local} often ascribe
$A(H\alpha) = 1.0$ to 1.5 mag (equivalent to $E(B-V) = 0.4$ to 0.6 mag)
to the individual galaxies
or the sample of galaxies studied in each case; see also
\cite{kennicutt83} and \cite{cohen01}.

There are many more surveys of emission lines in local star forming galaxies
using small apertures, which concentrate on the galactic nucleus, e.g.
\cite{mcquade95} and \cite{storchi95}.  We, however, rely on
total line fluxes, and these observations in general
cover too small a fraction of the galaxy as a whole to be useful
here.  \cite{rosa02} have assembled small aperture emission line
fluxes for
a set of 31 nearby star forming galaxies, combined them
with FIR fluxes from IRAS, and intercompared the
results of applying the various diagnostics.  With respect to SFR(FIR)
which measures the total SFR, they find, with no
extinction or inclination corrections, SFR(FIR) = 3.4 SFR(H$\alpha$),
with a large scatter (equivalent to
$\pm0.5$ dex) and SFR(FIR) = 6.0 SFR(3727), with a dispersion equivalent to
$\pm0.8$ dex.  Because we are using inclination corrections, our dispersions
should be and are somewhat smaller.
We will compare these relations to those we use later.

\section{The Rate of Star Formation Versus X-Ray Flux in the HDF
\label{section_hdf} }

As reviewed by \cite{kennicutt98}, there are many diagnostics
that can be used to infer star formation rates, ranging
from continuum emission in the UV, the far-IR and the radio
to recombination  and forbidden emission line strengths.
The systematic errors induced by dust 
in several of these relations were discussed
by \cite{bell02}, while \cite{buat02} discusses some of the
problems associated with dust extinction in the UV and the use
of UV fluxes, commonly used for galaxies with $z > 2.5$, where
the rest frame UV is redshifted into the  optical band.

Given the wavelength coverage of the
optical spectra from the CFGRS and the range
of redshift under consideration,
the SFR for the distant galaxies in our sample in the region of the HDF
can only be determined via
the flux in the 3727~\AA\ emission
line of [OII], which is one of the less robust of these diagnostics.
The strength of emission in this line is affected not only
by dust but also potentially by variation of O excitation 
and the O abundance  among the
galaxies.  The latter has been explored for local galaxies
by \cite{jansen01} and at intermediate redshifts
by \cite{cardiel03}.  All of these diagnostics are affected,
each weighting a slightly different mass range,
by the value of the initial mass function.

We measure the equivalent width in the 3727~\AA\ doublet, which
is not generally resolved in these spectra, in the observed
frame, then transform that to the rest frame.
SED parameters defined by the SED model of \cite{cohen01}
have been determined by \cite{cohen01}
for each of these galaxies
from multi-color broad band photometry
extending from U to K.  These
are used to interpolate within the 
set of observed broad band colors,
yielding the continuum flux at rest frame
3727~\AA.  Since the redshifts are known, the observed fluxes
in the emission line are  transformed into
emitted luminosity in this emission line for each galaxy in the sample.
We are forced to do this as the LRIS spectra are not fluxed;
multi-slit spectra cannot in general be fluxed due to 
differential slit losses between objects observed depending
on the accuracy of the slitmask alignment and of the
astrometry used to design the slitmasks.
The adopted method of calculating the rest frame equivalent
width of an emission line 
assumes either that the  line emission more or less
follows
the same spatial distribution as the background light
of the galaxy, or that the 1 arcsec wide slit used for
the spectroscopy includes most of the total light
of the galaxy.  The latter is a reasonable assumption at the
upper end of this redshift range; at
$z=0.8$, 1 arcsec corresponds to 7.3 kpc; the
full size of the galaxy is sampled in the direction of the
length of the slit.

The SFR is computed from the emitted flux in the [OII] line
using the calibration of \cite{kennicutt98}.  No corrections
for internal extinction within the galaxies are made.
The Galactic extinction in the direction of the HDF is very
small and is  ignored here.  We do, however, correct for 
inclination of the galaxies, which does not 
remove the reddening experienced by a face-on disk galaxy.
We measured rough axis ratios
from the images of the Flanking Field galaxies
obtained by the GOODS project \citep{giavalisco03}, and used the original HDF
images for those galaxies within the HDF itself.  Images taken
in the F814W filter, the reddest used in both these HST surveys,
were utilized when available.
In a few cases,
no HST image that was deep enough to indicate the shape of the outer
isophotes of the galaxy could be located; no inclination
correction was applied in such cases. 
We apply a correction of the form
$\Delta$(mag) = $\gamma {\rm{log}}(b/a)$.  We assume
$\gamma(3727) = 1.9$ mag, extrapolated from
the work of \cite{tully98}; this may be an underestimate of $\gamma$,
which is rarely determined at wavelengths bluer than B.  The SFR
thus determined is denoted SFR$^i$(3727).

There are two galaxies in the HDF sample with 
$z < 0.2$ for which
the equivalent width of H$\alpha$ is available
from the CFGRS spectra.  These, corrected for the inclusion
of [NII] in the blend and for inclination effects as described
above, yield SFR$^i$(H$\alpha$) in reasonable agreement with
SFR$^i$(3727).  These galaxies both are of low luminosity; the 
volume of the cone of the CFGRS redshift survey at such low $z$
is small.  One of these is the
object with the lowest star
formation in this HDF sample, F36332\_1134.
The diagnostic ratio of \cite{veilleux87} 
indicates that light from HII regions dominates
the optical emission lines.

The X-ray luminosity is taken directly from 
\cite{alexander03}; the observed flux over the full Chandra
bandpass is transformed into an emitted flux.  Many of
these sources are quite weak (see Figure~\ref{fig_xcounts}
and in some cases the X-ray
spectral index could not be determined from the Chandra data.
In a few cases, \cite{alexander03} tabulate only an upper limit to
the total flux over the full Chandra band, presumably because
of non-detection of the weakest sources in the higher frequency
part of the Chandra bandpass.
We therefore ignore any $k$-correction to the X-ray luminosity.

The resulting parameters for each galaxy 
in the CFGRS matched to within
1.5 arcsec to a Chandra point source
and which shows evidence of strong current
star formation from its optical spectrum are listed in Table~\ref{table_sfr}
and displayed in Figure~\ref{fig_hdf_inclincor}.  The final sample contains 
22 $\cal{E}$ galaxies and 16 $\cal{I}$ galaxies in the region of the HDF.
The median redshifts are $z=0.84$ for the latter and $z=0.49$ for
the latter; $\cal{I}$ galaxies are harder to identify at high redshift
\citep{cohen00}.
The two galaxies  that are discrepant
in Figure~\ref{fig_agn} and suspected of harboring AGNs
are marked in Figure~\ref{fig_hdf_inclincor}, as are the suspected
AGNs isolated from Figure~\ref{fig_xcounts}.

The range of SFR$^i$  at a fixed $L_X$ in Figure~\ref{fig_hdf_inclincor} 
is large, and is noticeably larger if the inclination corrections are
not used, while the uncertainties in the values of $L_X$ are small.
The differences in internal extinction
from galaxy to galaxy 
after being corrected to face-on is the dominant
contributor to the scatter in Figure~\ref{fig_hdf_inclincor}.
The effect of ignoring the X-ray k-correction is relatively small. 
The uncertainties in the 3727~\AA\ equivalent width measurements,
given in Table~\ref{table_sfr}, are in general small.
The conversion to an emitted
line flux may introduce errors at the 30\% level \citep{cohen01}.
The effect on the SFR$^i$ derived from the observed
luminosity in the 3727~\AA\ emission line
of an increase in the internal extinction
within a galaxy of
$E(B-V)=0.5$ mag is indicated by the horizontal arrow
at the lower right of the left panel of Figure~\ref{fig_hdf_inclincor}. 
Table~\ref{table_disp} summarizes the  expected
sources of uncertainty and the contributions of each.
Their sum, in quadrature, is in good agreement with the
dispersion seen in Figure~\ref{fig_hdf_inclincor}
of SFR$^i$(3727) for a given $L_X$.  This table offers a sobering reminder
of why SFR$^i$(H$\alpha$) is a more robust indicator of SFR than
is SFR$^i$(3727), but our options are limited.

Overall, aside from the suspected AGN, the HDF galaxies 
display a relationship between $L_X$ and SFR$^i$ in
Figure~\ref{fig_hdf_inclincor} which is highly
reminiscent of that of the local calibrators, with
larger scatter.  The only major
outlier among the $\cal{I}$ galaxies in the right panel
is a galaxy without an inclination correction.  The $\cal{E}$
galaxies, shown in the left panel, now show a tight relationship
at low $L_X$. At the highest X-ray
luminosities, the scatter is somewhat larger, while
the three most luminous among the $\cal{E}$ galaxies, and the
5 most luminous among the $\cal{I}$ galaxies are all suspected
AGN. It should
also be noted that 
two of these $\cal{E}$ galaxies with $L_X$ above 
that of the suspected AGN and with similar SFR,
F36348\_1628 and F36246\_1111, are not detected in the VLA radio surveys
of \cite{richards98} and \cite{richards00}.  Since they are both more distant
than the suspected AGN, that does not contradict the strong
suggestion from Figure~\ref{fig_xcounts} that these are AGN.

Table~\ref{table_median} lists the median properties of the samples
in the region of the HDF as well as those of the local comparison galaxies.
The second largest value for a given parameter among the galaxies
in a particular sample is listed in
Table~\ref{table_extreme}.  (We ignore the highest value to avoid
outliers.)
The SFR$^i$(3727) and SFR$^i$(H$\alpha$)
are given for a face-on galaxy
with no extinction correction.  

If we wish to determine the total SFR, we need to evaluate $\Delta(A)$,
the differential absorption between H$\alpha$ and 3727~\AA\
and also $A$(H$\alpha$).  For the reddening curve adopted
here, the former is given by
$\Delta(A) = 2.26E(B-V)$ mag, with $0 < \Delta(A)$, and
almost certainly $\Delta(A) < 1.7$ mag.  The second factor is
given by 2.67$E(B-V)$.  Then SFR(total) = $A$(H$\alpha$)SFR$^i$(H$\alpha$)
or $A$(H$\alpha$)$ \Delta(A)$SFR$^i$(3727).  If we assume a typical
$A_V$ of 1.5 mag for star forming galaxies, then
SFR(total) = 2.7 SFR$^i$(H$\alpha$) = 9.0 SFR$^i$(3727).
Recognizing their uncertainty is large,
we adopt these values hereafter as typical for both the local
and the HDF samples.  The comparable relations obtained for 31 local galaxies
by \cite{rosa02} have constants within 30\% of those given above
when either H$\alpha$ or the 3727~\AA\ emission line of [OII] is used
as the SFR diagnostic.

\section{Comparison of the SFR Between the HDF and Local Samples}

We have used the rest frame $K$ luminosity to infer the mass
of each galaxy.  We follow the procedure of \cite{cohen01},
where the SED parameters for each of the HDF galaxies
are determined to deduce their rest frame $K$ luminosity, a procedure
identical in principle to that described in \S\ref{section_hdf}
and used to determine the continuum
flux at rest frame 3727~\AA.  These
are the stellar masses; gas is not included.
Masses have been deduced for each of the local calibrators using
their integrated $K$ flux from 2MASS \citep{skrutskie97}.
The median mass of the local sample is $2.4\times10^{10} M$\subsun, 
comparable to that of the $\cal{E}$ galaxies in the HDF, while that
of the $\cal{I}$ galaxies is $\sim3$ times larger.  
As the range of galaxy masses is large, we compare the values of
SFR$^i/10^{11}M$\subsun\ for the HDF and for the local sample.
These, plotted as a function of mass, are shown in 
Figures~\ref{fig_hdfmass_sfr} and \ref{fig_localmass_sfr} respectively,
while the medians and the second highest values are given in
Tables~\ref{table_median} and Table~\ref{table_extreme}.
Recall that  for the
local sample we use SFR$^i$(H$\alpha$) while for
the HDF sample we use SFR$^i$(3727).  

The parameter SFR$^i/10^{11}M$\subsun\ spans a range of a factor
of $\sim100$ for the HDF sample, which we regard as an indication
of the rather broad galaxy spectral types used here, the lack
of any luminosity indicator in the optical spectra,
and our inability
to differentiate  $\cal{I}$ galaxies from $\cal{E}$ galaxies at $z \sim1$.
The $\cal{E}$ galaxy with the highest SFR$^i$/10$^{11}M$\subsun\ 
shown in Figure~\ref{fig_hdfmass_sfr} is the highest
redshift galaxy in our HDF sample, with $z=1.355$.  Its spectrum
has been checked; the [OII] line strength and its errors
appear valid.  This galaxy is not discrepant in the $L_X$ -- SFR
relationship of Figure~\ref{fig_hdf_inclincor}.  It 
appears to be
a genuine case of a high redshift galaxy with a very high
SFR for its mass.

With no correction for reddening, Table~\ref{table_median} shows that
the median SFR$^i/10^{11}M$\subsun\
for the $\cal{E}$ galaxies in the HDF 
is 2.6 times larger than that of the local galaxies, while that for
the $\cal{I}$ galaxies is larger by a factor of 1.9.  Thus,
irrespective of the exact value of the two correction factors 
discussed above, the total SFR for the HDF galaxies is considerably larger
than that of local galaxies.  We adopt the nominal correction factors
given above, with the additional support provided by the rough
agreement with the local sample of \cite{rosa02}.  The
median total SFR
is then 7 times larger in the HDF sample than
that of the local galaxies. The median
SFR per unit mass, SFR$^i/10^{11}M$\subsun, a measure of
the efficiency of SFR,  
for the HDF is $\sim1.7$ times that for the local sample with no 
extinction corrections.  Applying the nominal
extinction corrections suggests that the median SFR per unit mass is
5 times higher in the HDF.

We reject the suggestion that these differences arise from sample
selection,  such that only
galaxies with very high SFR can be detected at high redshift.
Sample selection in the form of incompleteness of the CFGRS 
is not believed to be a serious concern here.
We see many galaxies in the HDF with lower SFR (see 
Figure~\ref{fig_bighdf_lumz})
than those in the Chandra sample, at least for $z<1.1$.
In this context, we note that
the CFGRS could have detected emission in the 3727~\AA\ line
at least 5 times smaller than that observed for F36527\_1355
at $z=1.355$.

In local galaxies, a simple \cite{schmidt59} power law accurately relates
a galaxy's total SFR to its disk-averaged  gas surface density.
However, we do not know the gas density in these systems,
and information on possible mergers is not available either, so we
cannot attempt to isolate the 
additional factors beyond total stellar mass that undoubtedly influence 
the SFR in a particular galaxy.

In spite of these concerns, the similar form of the
relationship between $L_X$, SFR$^i$, $L$(radio) and $L$(FIR)
in HDF samples of varying sizes,
once cleaned of AGNs and taking reddening into consideration
(for SFR only), presented here, by \cite{grimm03}, and by \cite{garrett02}
suggest 
that the physical mechanism for star formation is similar among
all the galaxies discussed here,
and it is operating in more or less the same way, but 
galaxies of higher mass dominate the star formation in the
HDF, while lower mass galaxies dominate the star formation
locally, as was already pointed out by \cite{cowie96} and others.

\subsection{Constraints os the SFR At Still Higher Redshift}

We consider whether the present SFR in each galaxy in the HDF sample
is consistent with its age as inferred from its
redshift $z$, or whether the time averaged SFR had to be even higher
in the past to have formed the observed mass of stars.
This provides a window into the behavior of SFR with $z$ \citep{madau98}
in a redshift regime 
where observations are more difficult.
Figure~\ref{fig_sfr_time} shows SFR$^i$(3727) for the HDF sample
as a function of galaxy mass.  The solid line indicates
the accumulated stellar mass of a galaxy after 6$\times10^{9}$ yr
(the age of a $z\sim1$ galaxy) for
a SFR constant with time.  If a galaxy is located to the right of the line,
there has been insufficient time to produce the observed stellar content
of the galaxy assuming SFR constant with time.  In such a case,
the time averaged SFR must have been higher in the past (i.e. presumably
at $z$ between 1 and 2, as the total age at $z>2$ is very small,
$\lesssim$1 Gyr).

The symbol size in this figure increases with redshift, so we
check whether the galaxies with $z > 0.7$ are
systematically to the right of the line.
The $\cal{E}$ galaxies are distributed fairly close to this line,
while the $\cal{I}$ galaxies tend to be to the right of it.
Their median redshift is half that of the $\cal{E}$ galaxies
(Table~\ref{table_median}), 
but the dashed line in the right panel of the figure indicates
the mass expected for constant SFR and an age of 9 Gyr;  the 
$\cal{I}$ galaxies
still tend to lie to the right of that.

For the $\cal{E}$ galaxies in the HDF,
only one high redshift galaxy appears to be
more than a factor of three from the line.
This case,
and the smaller offsets to the right of the line of the other
$z\sim1$ galaxies, 
can easily be accommodated by the factors $\Delta(A)$ and $A$(H$\alpha$), both
assumed to be unity here, as well as by a few missing inclination
corrections.  So there is no need for a substantially
higher SFR at $z>1$ to produce the stars seen in galaxies at $z \sim1$.
The current SFR at $z \sim1$ is consistent with the time averaged SFR
for $z>1$ for the $\cal{E}$ galaxies.  It is possible, but not 
demonstrated here,
that the absorption corrections $A($H$\alpha)\Delta(A)$ 
can also produce consistency
for the $\cal{I}$ galaxies without requiring a higher time
averaged SFR in the past.  Further tests of this once more
data become available will be interesting.

\section{The Origin of the X-Ray Emission from These Galaxies}

We now consider which star forming
galaxies within the HDF show X-ray emission.
\cite{cohen03} presents rest frame equivalent widths for the
3727~\AA\ emission line of [OII] in a sample of 256 galaxies
from the CFGRS in the region of the HDF.  Figure~\ref{fig_bighdf_z}
shows these as a function of redshift, with the X-ray luminous objects
of the present sample
indicated by larger symbols.  We see that the X-ray luminous
objects are not in general those galaxies at a given $z$ with the
largest equivalent widths, but are mixed through the entire
range of equivalent width, with a concentration towards the
lower values.  However, the equivalent width is not
a measure of the
emitted luminosity in the line.  When we plot instead
the emitted luminosity in the [OII] emission line at 3727~\AA\
(see Figure~\ref{fig_bighdf_lumz}) versus redshift, the situation
is somewhat clearer.  The X-ray luminous galaxies are among the
most luminous in the full HDF sample of the CFGRS
in this emission line, i.e. they have among the
highest SFR at each $z$, but having a very high SFR does
not guarantee that a galaxy will be detected in the Chandra images.
We cannot at present isolate the additional factors beyond the mass 
of a galaxy that might influence its SFR at $z \sim1$.

However, Figure~\ref{fig_xcounts} offers some possible clues.
We suggest that the small number of
discrepant galaxies in this figure all have weak nuclear AGNs.
We suggest that the normal star forming galaxies are those sprinkled
along the bottom of this figure.  These all have very low full-band
Chandra counts,
almost all less than 20 full-band counts.  This is right at the 
detection limit of even the 2 Ms Chandra image \citep{alexander03}.
The small variations in inclination angle and reddening mean that
those galaxies with the maximum detected 3727~\AA\ line flux
at any given redshift (i.e. the upper envelope of
the distribution in Figure~\ref{fig_bighdf_lumz})
can only suggest, but not define, the set of galaxies with the highest
total SFR.
The lack of Chandra detection of galaxies apparently with slightly higher or
comparable SFR to those detected becomes a matter of chance, the
inclination or the reddening. The 
spatial position of the galaxy within the Chandra field,
in particular its distance from the center of the
Chandra field (where detection limits are higher), may also
play a role.  With this
hypothesis to explain why some star forming galaxies we would expect to have been
detected by Chandra are not,
the $L_X-SFR$ relation among the star forming galaxies in the region
of the HDF becomes indistinguishable from that of local galaxies, and
$L_X \propto SFR$.

\section{Summary}

There have been many efforts to determine the SFR at high redshift.
The evolution with redshift of the rest frame UV luminosity density
was used by, among others, \cite{lilly96}, \cite{cohen02} and \cite{wilson02}
to study the evolution of the SFR with $z$, while \cite{hogg98}
used the evolution of the [OII] luminosity density for the same purpose.
All agree on a strong increase in these SFR diagnostics between
the local Universe and $z\sim1$.  In the present analysis 
we see this manifested through the
emission line strengths of individual galaxies;
the very
high emitted luminosity in the 3727~\AA\ line of
[OII] and the inferred SFR found among the most extreme 
galaxies in the HDF sample (see Tables~\ref{table_sfr} and
\ref{table_median})
surpass anything seen in the local Universe.  The enhancement
in SFR/unit galaxy mass is also substantial. 

\begin{figure}
\epsscale{0.9}
\plotone{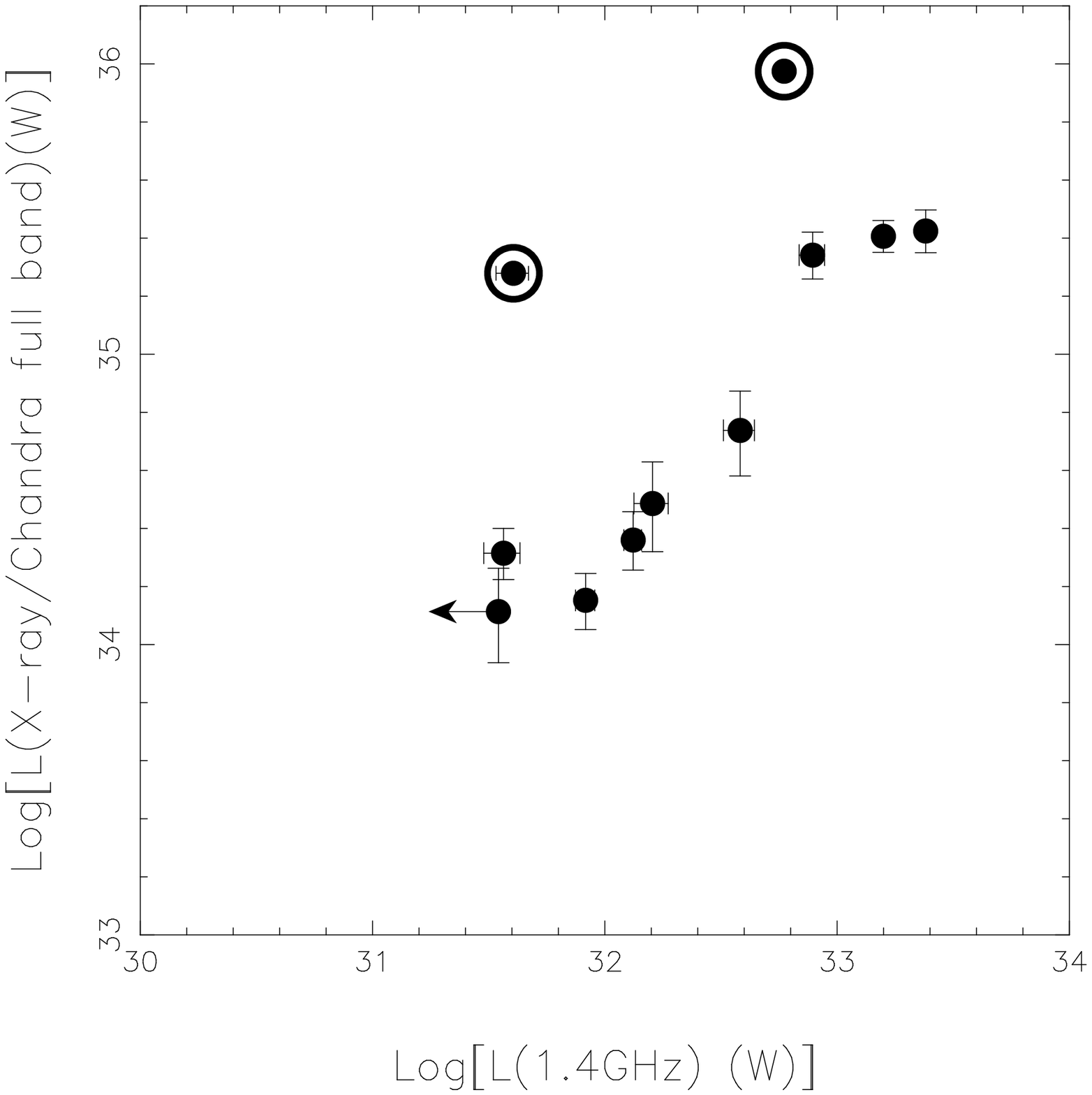}
\caption[]{The total X-ray luminosity over the full  Chandra band 
is shown as a function of the
radio luminosity at 1.4 GHz from \cite{richards98} and \cite{richards00}
for a sample of 11
star forming galaxies in the region of the HDF which appear as point
sources in the Chandra catalog of \cite{alexander03}, are detected with
the VLA, have redshifts from the CFGRS, and meet several other
criteria described in the text. The two suspected AGNs are circled.
\label{fig_agn}}
\end{figure}

\begin{figure}
\epsscale{0.3}
\plotone{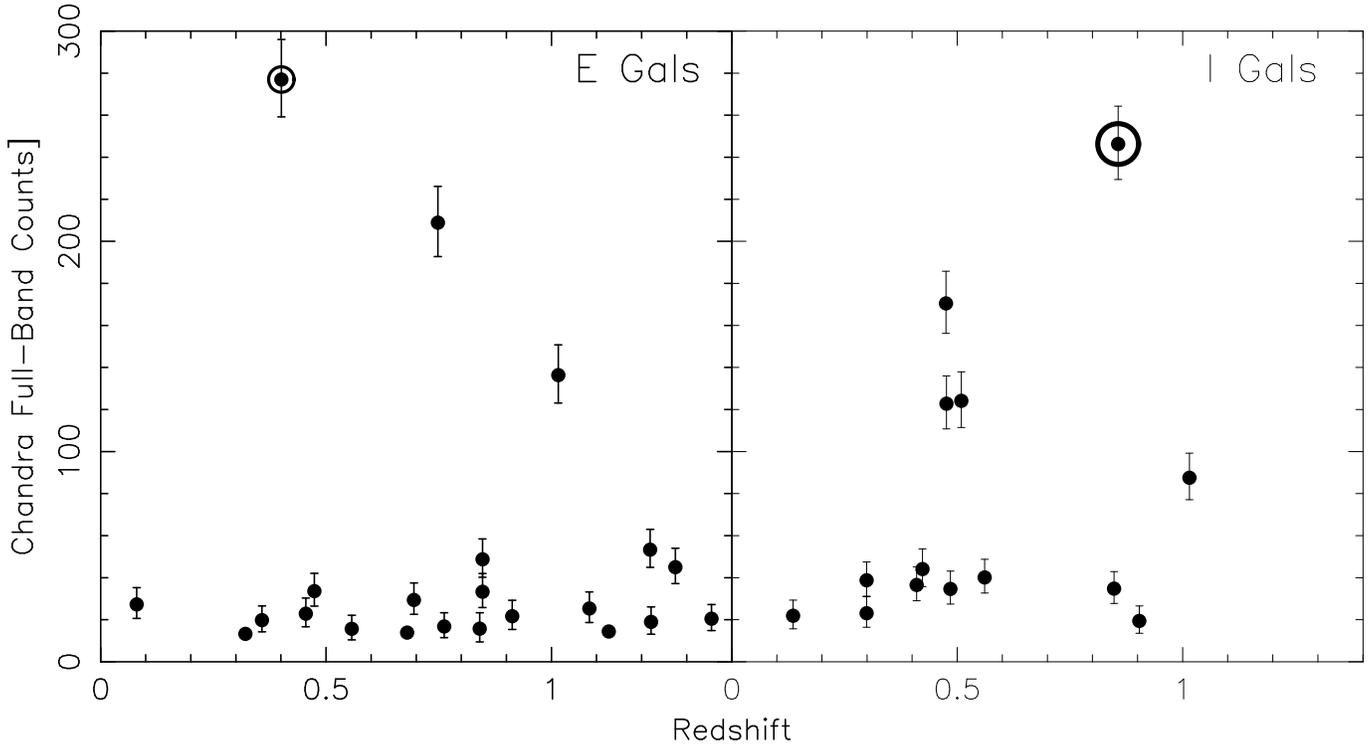}
\caption[]{The detected counts over the full Chandra band 
are shown as a function of the galaxy redshift
for 22 galaxies with very strong emission lines (left panel)
and 16 galaxies with moderately strong emission lines (right panel) from 
the overlap
of the CFGRS and the Chandra point source list in the region of the HDF-N.
Two of the $\cal{I}$ galaxies have have
more than 1000 detected Chandra full band counts, and hence do not appear
in this plot.
\label{fig_xcounts}}
\end{figure}

\begin{figure}
\epsscale{0.9}
\plotone{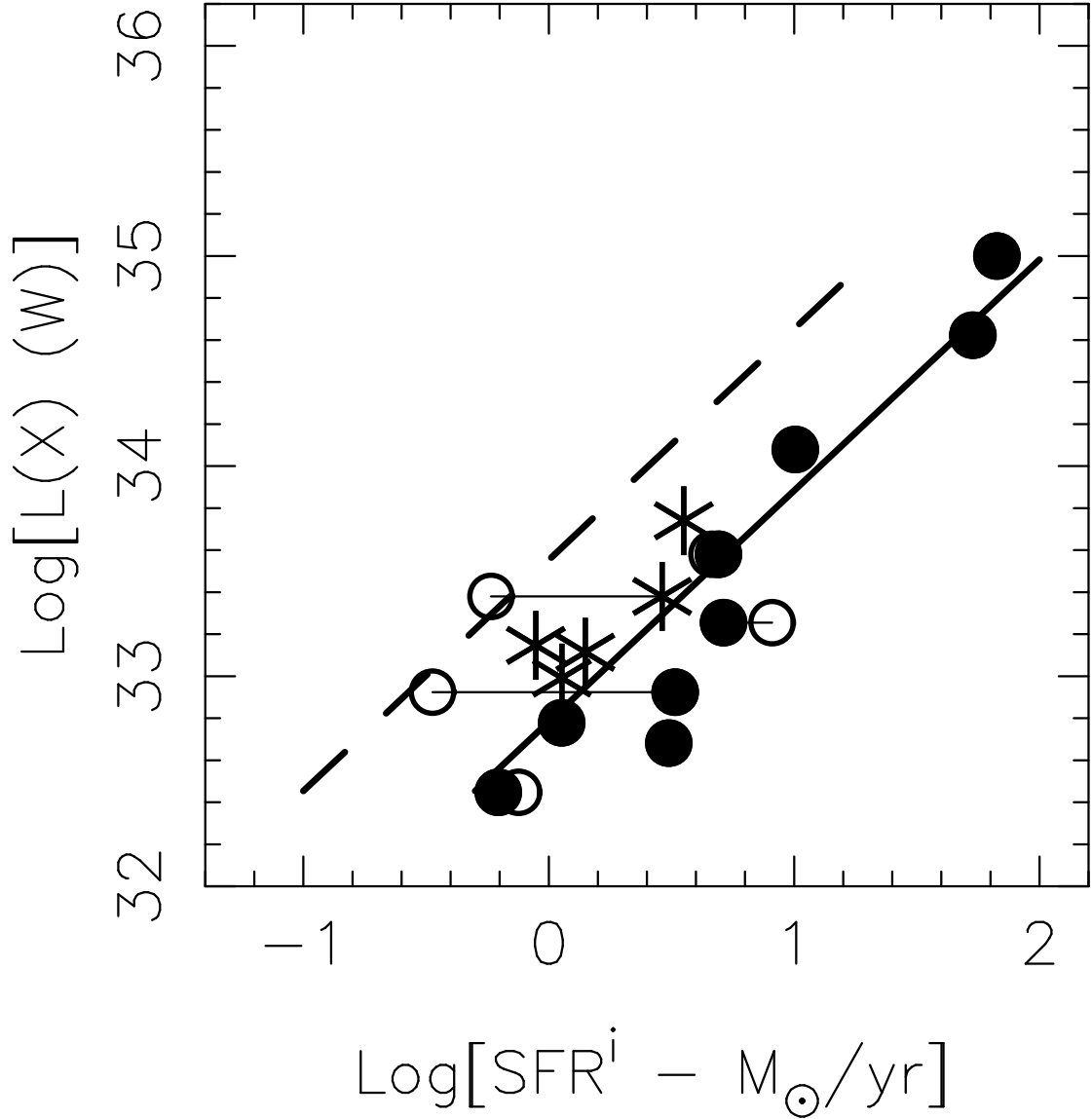}
\caption[]{The total X-ray luminosity over the full  Chandra band 
is shown as a function of the SFR deduced from the H$\alpha$ flux
corrected for inclination effects for the sample of 14 local
calibrating star forming galaxies.  The filled circles represent
SFR derived from H$\alpha$ luminosities for galaxies
with $(B-V)_0 > 0.6$ while the stars indicate the
position of the redder star forming galaxies.
The open circles
use the 3727~\AA\ emission line of [OII] as a diagnostic.  Thin
solid horizontal lines connect galaxies with measurements using
both diagnostic lines.   
The solid line denotes the
least squares fit.  The dashed line is offset in SFR$^i$
from that by $-0.7$ dex, corresponding to the maximum offset
seen for SFR$^i$(3727) with respect to SFR$^i$(H$\alpha$).
\label{fig_local_inclincor}}
\end{figure}

\begin{figure}
\epsscale{0.3}
\plotone{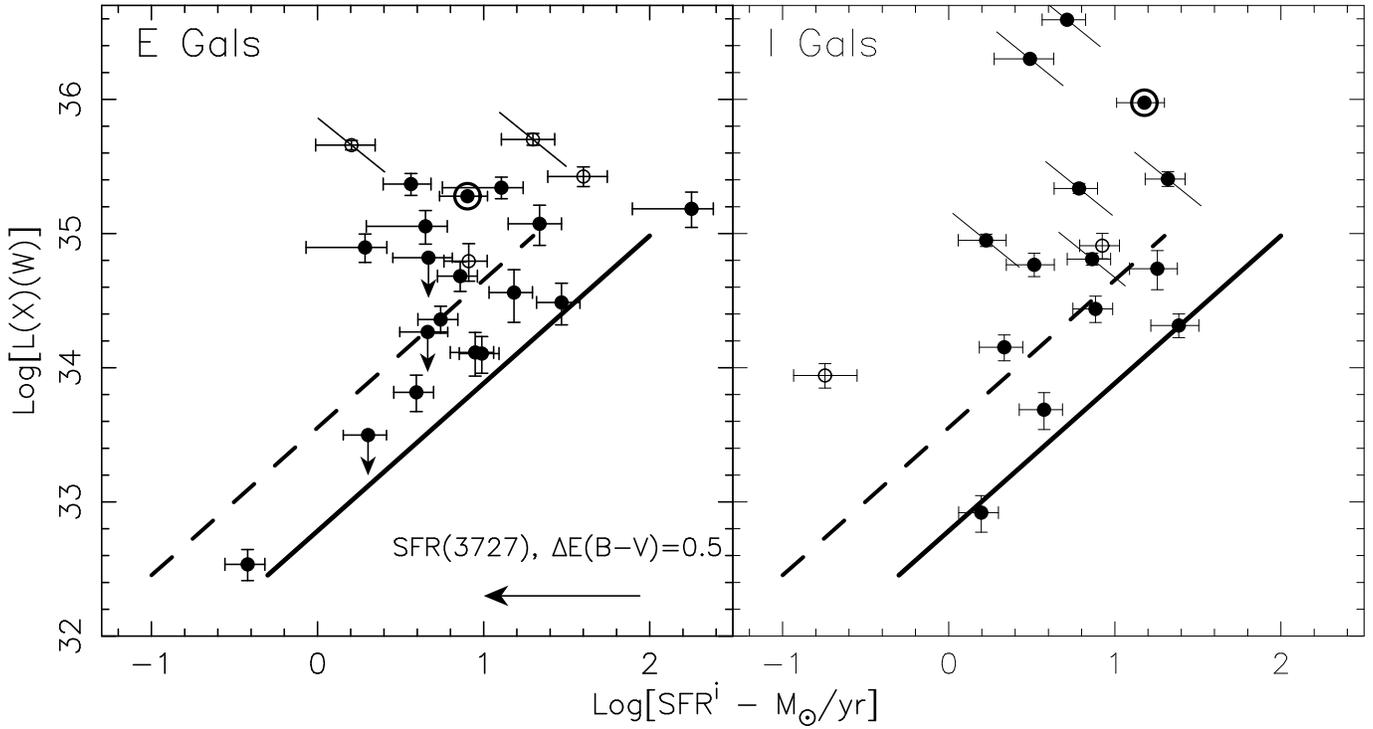}
\caption[]{The total X-ray luminosity over the full  Chandra band 
is shown as a function of the SFR deduced from the luminosity in the
3727~\AA\ emission line of [OII] 
corrected for inclination effects for 
22 galaxies with very strong emission lines (left panel)
and 16 galaxies with moderately strong emission lines (right panel) from 
the overlap
of the CFGRS and the Chandra point source list in the region of the HDF-N.
Open circles indicate galaxies lacking inclination corrections. 
The two suspected AGNs isolated from Figure~\ref{fig_agn} are circled.
The additional suspected AGNs isolated from Figure~\ref{fig_xcounts}
are marked with short diagonal lines.
The least squares linear fit to the local sample when
H$\alpha$ is used as the diagnostic for SFR is indicated as the
thick solid line, while the dashed line denotes that expected
from the 3727~\AA\ emission line of [OII].  
The horizontal arrow near the bottom of the left panel
denotes the decrease in SFR$^i$(3727) expected if
$E(B-V)$ is increased by 0.5 mag.
\label{fig_hdf_inclincor}}
\end{figure}

\begin{figure}
\epsscale{0.9}
\plotone{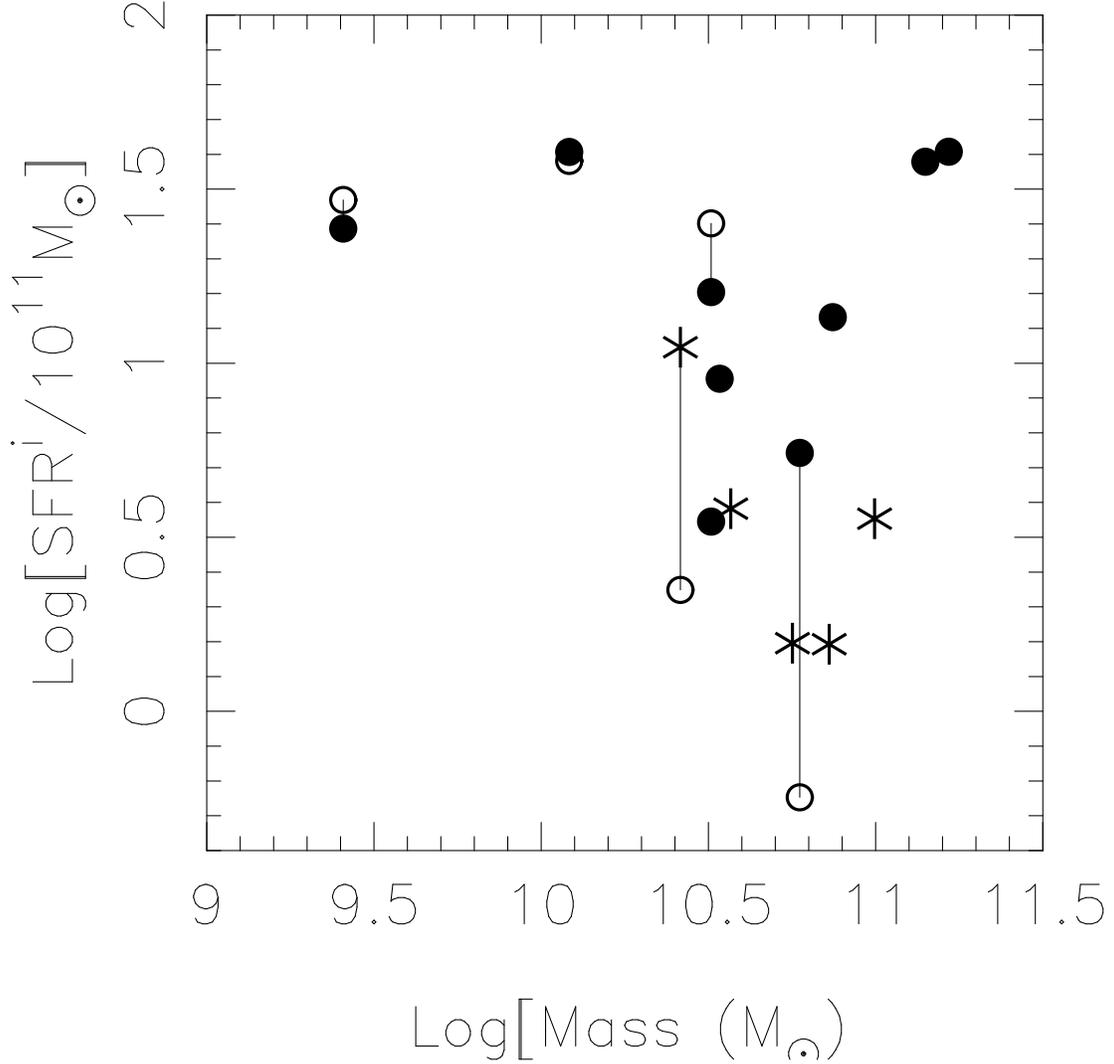}
\caption[]{SFR$^i$(H$\alpha$)/10$^{11}M$\subsun\ is shown as a 
function of galaxy mass for the 14 galaxies in the local sample.
The filled circles represent
SFR derived from H$\alpha$ luminosities for galaxies
with $(B-V)_0 < 0.6$ while the stars indicate the
position of the redder star forming galaxies.
Open circles denote the same, using the 3727~\AA\ line of [OII]
as a diagnostic of SFR.  Vertical lines connect the H$\alpha$ and [OII] 
measurements, when both are available.
\label{fig_localmass_sfr}}
\end{figure}

\begin{figure}
\epsscale{1.0}
\plotone{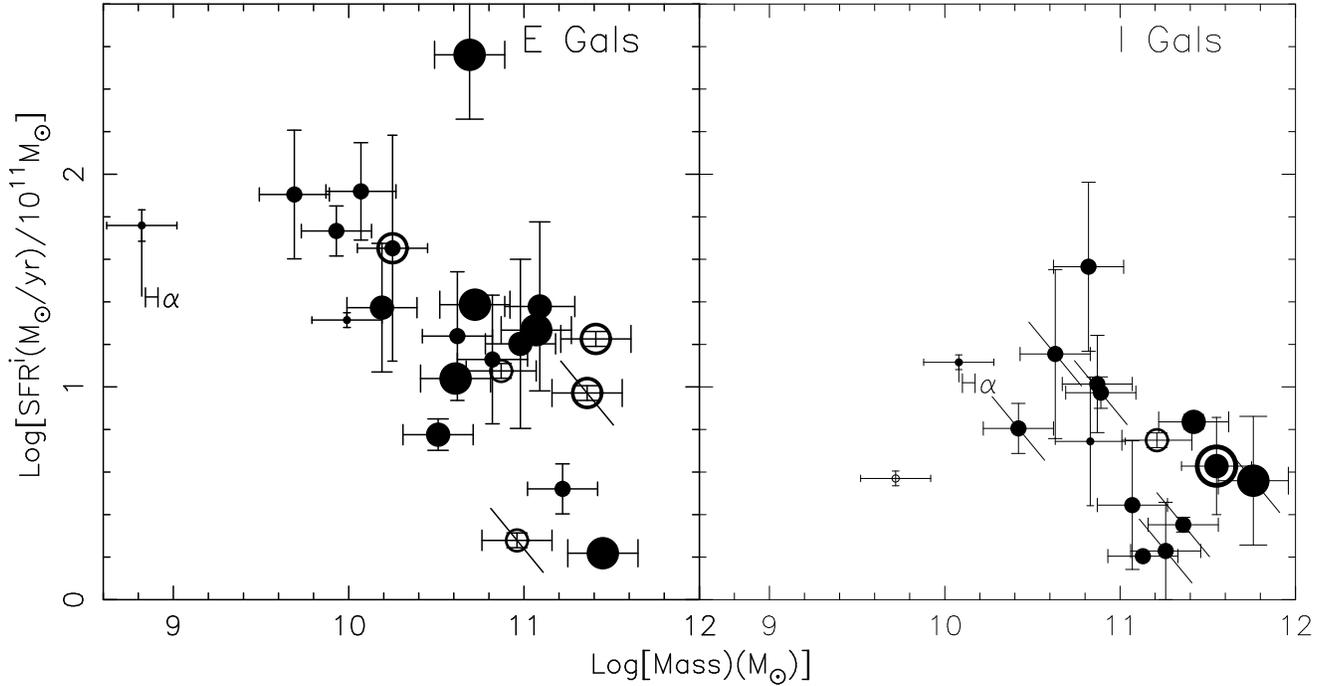}
\caption[]{The  SFR$^i$/10$^{11}M$\subsun 
is shown as a function of the galaxy mass 
for 22 galaxies with very strong emission lines (left panel)
and 16 galaxies with moderately strong emission lines (right panel) from the overlap
of the CFGRS and the Chandra point source list in the region of the HDF.
Galaxies without inclination corrections are shown as open circles.
The symbol size indicates the redshift of the galaxy; the smallest
symbols denote galaxies with $z < 0.35$, intermediate sized symbols
are used for $0.35 < z < 0.7$ and larger symbols for $0.7 < z > 1.0$.
Galaxies with $z > 1.0$ have the largest symbols.
The two AGNs isolated from Figure~\ref{fig_agn} are circled, while the
additional suspected AGNs isolated from Figure~\ref{fig_xcounts}
are indicated by diagonal lines.  The
position of each of the two galaxies with measured SFR(H$\alpha$) are indicated.
\label{fig_hdfmass_sfr}}
\end{figure}

\begin{figure}
\epsscale{0.9}
\plotone{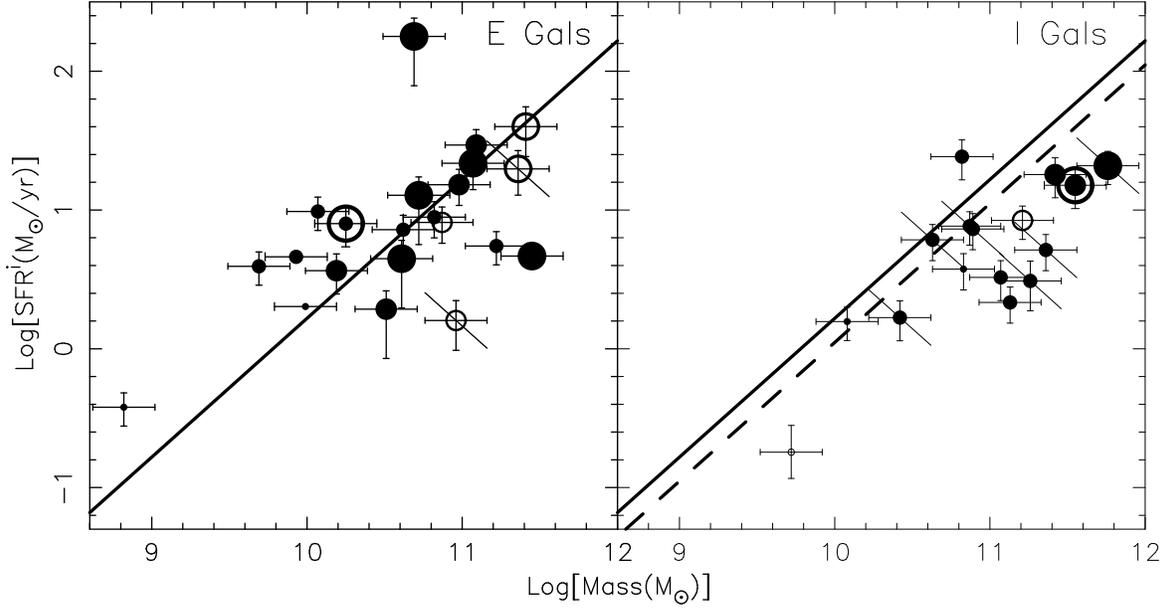}
\caption[]{SFR$^i$(3727) is shown for the
sample of X-ray emitting galaxies in the region of the HDF as a function
of galaxy stellar mass.
The right panel shows the 16 $\cal{I}$ galaxies while the left
panel shows the 22 $\cal{E}$ galaxies.
The symbol size indicates the redshift of the galaxy; the smallest
symbols denote galaxies with $z < 0.35$, intermediate sized symbols
are used for $0.35 < z < 0.7$, large symbols for $0.7 < z < 1.0$.
The largest symbols denote galaxies with $z > 1$.
Open circles indicate galaxies lacking measured axis ratios.
The two AGNs isolated from Figure~\ref{fig_agn} are circled, while the
additional suspected AGNs isolated from Figure~\ref{fig_xcounts}
are indicated by diagonal lines.
The line denotes the mass in stars
achieved as  function of SFR$^i$ after 6$\times10^9$ yr assuming
constant SFR with time.  The dashed line in the right panel
only is for an elapsed time of 9 Gyr.
\label{fig_sfr_time}}
\end{figure}

\begin{figure}
\epsscale{0.9}
\plotone{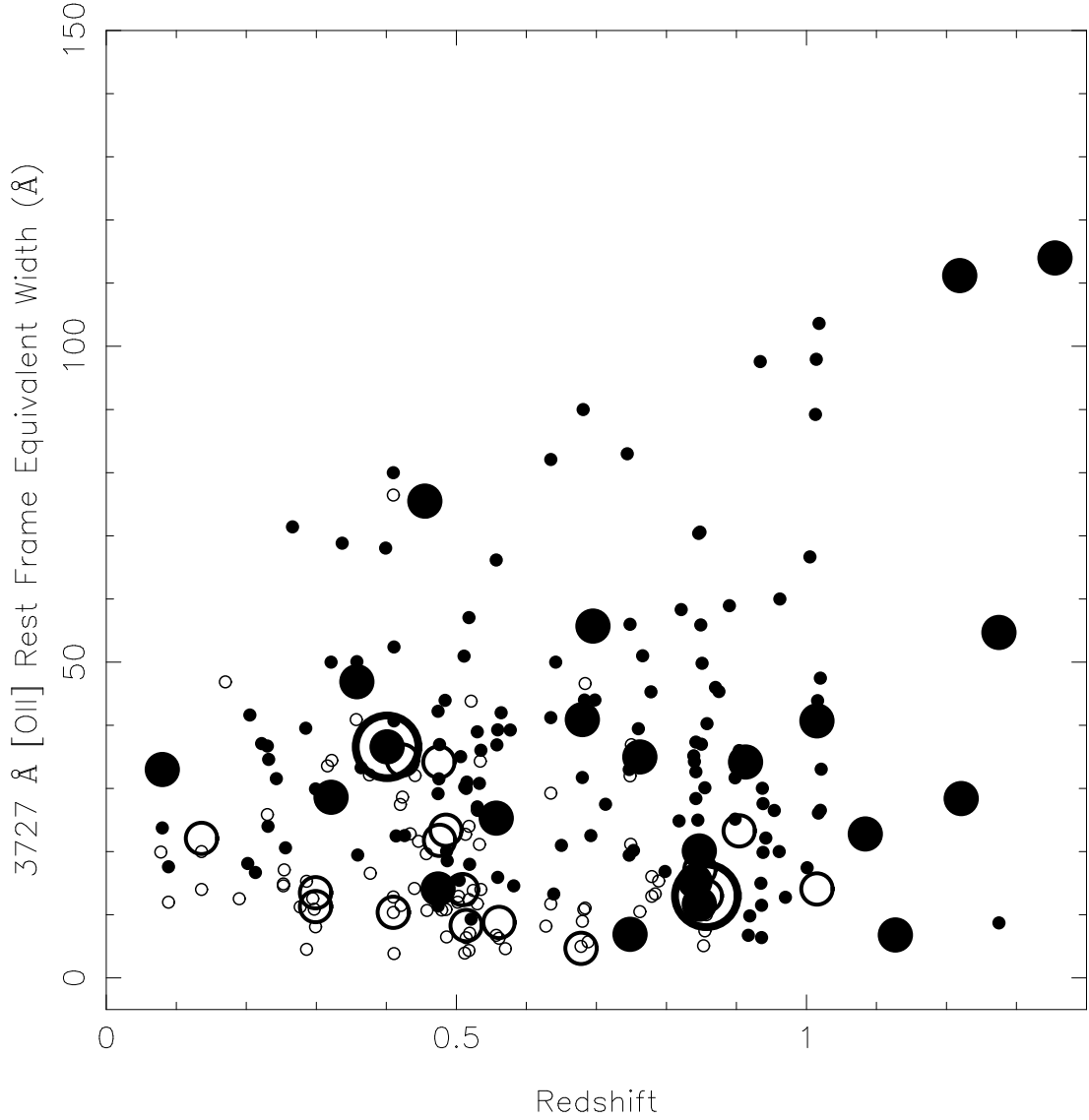}
\caption[]{The rest frame equivalent width of the 3727~\AA\ emission
line of [OII] is shown for a sample of $\sim$200 star-forming galaxies in 
the region of the HDF from \cite{cohen03}, ignoring a few broad lined
AGNs. 
Open circles indicate $\cal{I}$ galaxies while filled circles indicate
$\cal{E}$ galaxies.  
The X-ray emitting galaxies are marked by larger symbols.  
The two suspected AGNs are circled.  
\label{fig_bighdf_z}}
\end{figure}

\begin{figure}
\epsscale{0.9}
\plotone{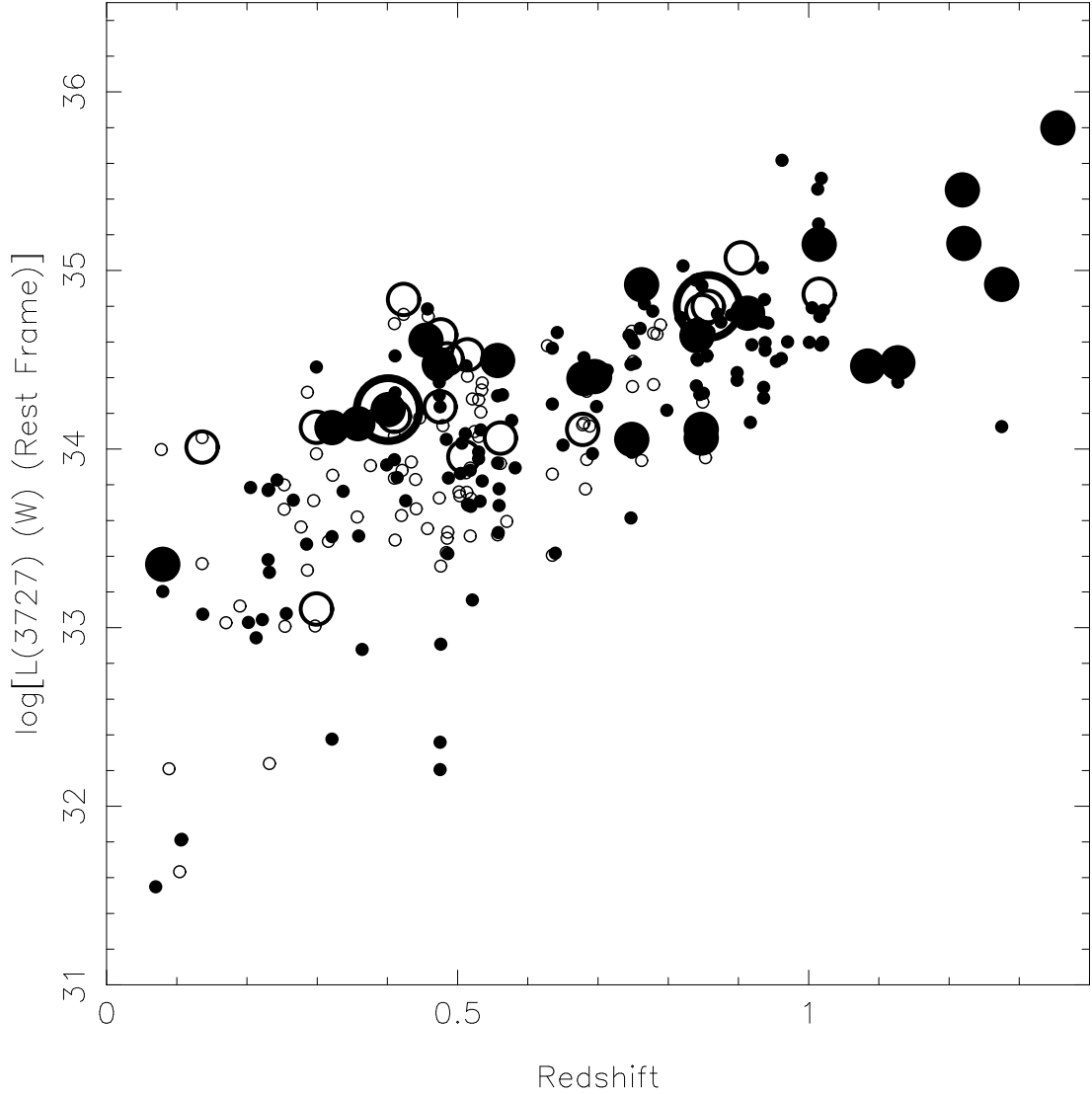}
\caption[]{The emitted luminosity in the rest frame 
in the 3727~\AA\ emission
line of [OII] is shown for a sample of $\sim$200 star-forming galaxies in 
the region of the HDF from \cite{cohen03}, ignoring a few broad lined
AGNs. 
Open circles indicate $\cal{I}$ galaxies while filled circles indicate
$\cal{E}$ galaxies.  
The X-ray emitting galaxies are marked by larger symbols.  The two suspected
AGNs are circled.  No inclination corrections have been made.
\label{fig_bighdf_lumz}}
\end{figure}

A comparison of the X-ray, radio and optical data 
suggests
(Figures~\ref{fig_agn} and \ref{fig_xcounts}), in agreement
with
evidence from other surveys, that 
$\sim$20\% of the X-ray sources contain weak AGNs,
even though these galaxies (ignoring the two broad lined AGNs
that have been deleted from the HDF sample)
appear from optical spectra of their integrated light 
to be ``normal''.  Aperture effects appear to limit our ability to
detect AGN from such spectra in high redshift samples
as suggested by \cite{moran02}.
These AGN
produce more than 90\% of the detected Chandra flux arising
from objects in the redshift regime $0.4 < z < 1.1$.

Because we are using the 3727~\AA\ emission line of [OII] as our diagnostic
of SFR, reddening is a serious concern, and we found that the
introduction of inclination corrections significantly reduced
the scatter in the $L_X-SFR$ relation.
This is the first detection to our knowledge of inclination effects
in such distant galaxies.  It is very likely that inclination effects
also contribute to the distribution of SEDs among distant star forming
galaxies.  They probably cause a significant part of the spread observed
in the SED parameters $\alpha_{UV}$ and $T$(sBB) determined by
\cite{cohen01}
for the set of $\cal{I}$ and for the set $\cal{E}$ galaxies in the
region of the HDF.

Once the reddening corrections are incorporated
and the AGNs eliminated, we suggest that the
$L_X-SFR$ relation for these distant galaxies is essentially identical
to that prevailing in local star forming galaxies.  The $L_X$-radio
relation, presented  here and, for a smaller sample, by \cite{grimm03}, and 
the $L_X-FIR$ relation of \cite{garrett02} for the HDF all
agree with those of local star forming galaxies.  This suggests
that the physical mechanisms responsible for X-ray
emission in these distant galaxies are the same as those
that act locally. 

High observed flux in the 3727~\AA\ line of [OII]
is a necessary but not a sufficient condition for detection by Chandra. 
The issue of
which particular galaxies are detected by Chandra is resolved by
realizing that the normal star forming galaxies in our sample
lie essentially at or only slightly above the Chandra detection limit.
Inclination, reddening, etc which affect the transformation from [OII]
emission line flux to total SFR
can lead to a particular galaxy being detected
by Chandra, whereas another with similar SFR$^i$(3727) at the same redshift
may not be detected.

Many optical surveys of distant galaxies are underway
which plan to use 3727~\AA\ emission as a diagnostic for SFR
and for metallicity.
In qualitative
terms, if the problems of internal reddening and 
potential AGN contribution can be handled,
the emitted luminosity in the
3727~\AA\ emission line does appear to be a reasonable indicator
of the SFR among these distant galaxies, but this work illustrates
the difficulties associated with this emission line, where reddening
is such a serious concern.

Once larger samples of distant galaxies with high precision multi-wavelength 
data permitting the use of multiple diagnostics
for their star formation rates become available, one might be able
to explore the issue of
the constancy
of the initial mass function between the local and the distant
Universe.  The comparison 
we have carried out of the expected stellar mass assuming constant
SFR with the actual galaxy stellar mass at $z\sim1$
suggests that no further rise in time averaged SFR is necessary 
at earlier times.  Extending this to a larger sample with better
data will be of considerable interest as it  probes 
the time averaged SFR at a key epoch of the formation of galaxies.

\acknowledgements

The entire Keck/HIRES user community owes a huge debt
to Jerry Nelson, Gerry Smith, Bev Oke, and many other people who have
worked to make the Keck Telescope and LRIS a reality and to
operate and maintain the Keck Observatory.  
We are grateful to the W. M. Keck Foundation for the vision to 
fund the construction of the W. M. Keck Observatory. 
The author extends special thanks to those of Hawaiian ancestry
on whose sacred mountain we are privileged to be guests. 
Without their generous hospitality, none of the observations presented
herein would have been possible. 

I am grateful to R. Sunyaev for helpful conversations and to
my collaborators in the HDF-N redshift survey, in particular
A.Phillips and Len Cowie,
who provided in digital form spectra of 
several galaxies in the HDF.  I thank the referee for
helpful suggestions.

This publication makes use of data products from the Two Micron
All Sky Survey, which is a joint project of the University of
Massachusetts and the Infrared Processing and Analysis Center/
California Institute of Technology, funded by the National
Aeronautics and Space Administration and by the National Science
Foundation.
This research has made use of the NASA/IPAC Extragalactic Database (NED)
which is operated by the Jet Propulsion Laboratory, California
Institute of Technology, under contract with the National Aeronautics
and Space Administration.

The extragalactic work of the author is not supported by any
federal agency.

\clearpage

\begin{deluxetable}{lrr rrrr}
\tablenum{1}
\tablewidth{0pt}
\small
\tablecaption{Local Calibrating Galaxies \label{table_local}}
\tablehead{ 
\colhead{ID} &  \colhead{log[$L_X$]} 
& \colhead{log(Mass)}  & \colhead{log[SFR$^i$(H$\alpha$)}
& \colhead{Log[SFR$^i$(3727)} &
\colhead{Refs.} & \colhead{Refs} \\
\colhead{} &  \colhead{(log(W))} & \colhead{(log[$M$\subsun])} &
\colhead{$/10^{11}M$\subsun ($M$\subsun/yr)]} & \colhead{($M$\subsun/yr)]} &
\colhead{(X-ray)} & \colhead{(Opt.)} } 
\startdata
NGC~253 & 33.11 & 10.57 & 0.58 & \nodata & 1 & $i$ \\
NGC~628 & 32.78 & 10.51 & 0.55 & \nodata & 1 & $i$ \\
NGC~891 & 32.99 & 10.86 & 0.19 & \nodata & 9 & $i$ \\
NGC~1614 & 34.62 & 11.15 & 1.58 & \nodata  & 9 & $ii$ \\
NGC~3031 (M81) & 33.15 & 10.75 & 0.20 & \nodata  & 1 & $i$ \\
NGC~3034 (M82) & 33.38 & 10.42 & 1.05 & $-$0.24 & 2 & $i$ \\
NGC~3256 &  35.00 & 11.22 & 1.61  & \nodata & 3,10 & $ii$ \\
NGC~3310 & 33.58 & 10.08 & 1.61 & 0.67   & 1 & $i$ \\
NGC~4038\tablenotemark{c} &  34.08 & 10.87 & 1.13 & \nodata & 4 & $i,iii$ \\
NGC~4321 (M100) &   33.74 & 11.00 & 0.55 & \nodata  & 5 & $i$ \\
NGC~4449 & 32.45 & 9.41 & 1.58 & $-$0.12  & 1,6 & $i$ \\
NGC~4631 & 33.26  & 10.51 & 1.20 & 0.49  & 1 & $i$ \\
NGC~5194 (M51) &  32.92 & 10.77 & 0.74 & $-$0.47  & 1,7 & $i$ \\
NGC~5236 (M83) & 32.68 & 10.53 & 0.96 & \nodata  & 8 & $i$ \\
\enddata
\tablenotetext{a}{1. \cite{fabbiano92},
2. \cite{ptak99}, 3. \cite{lira02}, 
4. \cite{fabbiano01}, 5. \cite{immler98}, 6. \cite{vogler97},
7. \cite{terashima01},
8. \cite{okada97}, 9.\cite{ueda01}, 10. \cite{moran99}.}
\tablenotetext{b}{$i$. \cite{bell01}, $ii$. \cite{calzetti94}, 
\cite{calzetti95}, $iii$. \cite{young96} }
\tablenotetext{c}{Also known as the Antennae galaxy.}
\end{deluxetable}

\begin{deluxetable}{lll}
\tablenum{2}
\tablewidth{0pt}
\small
\tablecaption{Contributions to the Scatter of SFR$^i$ versus $L_X$ \label{table_disp}}
\tablehead{ 
\colhead{Concern} &  \colhead{$\sigma$log[SFR$^i$]} 
& \colhead{$\sigma$log[SFR$^i$]} \\
\colhead{} &  \colhead{(3727) (dex)} & \colhead{(H$\alpha$) (dex)} } 
\startdata
Optical: \\
${\sigma}[E(B-V)] = 0.2$ mag & 0.4 & 0.2 \\
O abundance & 0.1\tablenotemark{a} & 0.0 \\
O excitation & 0.2 & 0.0 \\
Optical flux unc. & 0.15 & 0.1 \\
Optical ap. corr. & 0.05\tablenotemark{b} & \nodata \\
Inclination correc. unc. & 0.1 & 0.05 \\
X-ray: \\
No X-ray $k$-correc. & 0.1 & 0.1 \\
Conv. to Chandra\tablenotemark{c} & 0.05 & 0.05 \\
\hline
Predicted total $\sigma$\tablenotemark{d} & 0.51 & 0.26 \\
Observed $\sigma$ & \nodata & 0.31 \\
\enddata
\tablenotetext{a}{Calculated from the fit of 3727/H$\alpha$ 
to a sample of local galaxies of known O/H by \cite{jansen01}.} 
\tablenotetext{b}{The spectra are generally obtained through a slit that
is too small to pass all the light from the galaxy.  See the text for
how this is taken into account.  This term is the uncertainty in the correction.}
\tablenotetext{c}{Conversion between ASCA, Einstein and Chandra full-band fluxes.}
\tablenotetext{d}{This is the sum in quadrature of the terms listed above.}
\end{deluxetable}

\clearpage

\begin{deluxetable}{lrccr}
\tablenum{3}
\tablewidth{0pt}
\small
\tablecaption{Star Formation Rates for Galaxies in the HDF-N \label{table_sfr}}
\tablehead{ 
\colhead{ID} &  \colhead{$z$} & \colhead{$W_{\lambda}$(3727)\tablenotemark{a}} &
\colhead{log[SFR$^i$(3727)} & \colhead{Radio/VLA\tablenotemark{b}} \\
\colhead{} &  \colhead{} &
\colhead{(\AA)} & \colhead{(log[$M$\subsun/yr])} }
\startdata
$\cal{E}$ Galaxies \\
F36194\_1252 &   0.474 & 14.1$\pm1.5$ &  0.74($+0.10,-0.14)$ & Y \\
F36199\_1251 &   0.695 & 55.7$\pm5.0$ &  0.86($+0.10,-0.14$) \\
F36211\_1208 &   0.841 & 15.3$\pm2.0$ &  1.18($+0.11,-0.15$) \\
F36246\_1111\tablenotemark{c} 
    &   0.748 &  6.9$\pm2.0$ &  0.20($+0.14,-0.21$) \\
F36273\_1258 &   1.221 & 28.4$\pm7.1$ &  1.34($+0.13,-0.19$) \\
F36312\_1236 &   0.455 & 75.5$\pm7.6$ &  0.99($+0.10,-0.13$) \\
F36332\_1134 &   0.080 & 33.0$\pm3.0$ &  $-0.42\pm0.12$  \\
F36336\_1005\tablenotemark{c}
      &   1.015 & 40.7$\pm10.0$ &    1.30($+0.13,-0.19$) \\
F36346\_1241\tablenotemark{c}
      &   1.219 & 111.2$\pm33.3$ &  1.60($+0.14,-0.21$) & Y \\
F36348\_1628 &   0.847 & 11.7$\pm2.2$ &    0.56($+0.12,-0.17$) \\
F36363\_1320 &   0.680 & 40.9$\pm8.2$ &    0.66$\pm0.12$ \\
F36389\_1257 &   1.127 & 6.8$\pm2.0$ &     0.67($+0.14,-0.23$) \\
F36440\_1250 &   0.557 & 25.3$\pm7.6$ &    0.95($+0.11,-0.15$) & Y \\
F36470\_1237 &   0.321 & 28.6$\pm4.3$ &    0.30($+0.11,-0.15$) \\
F36517\_1220 &   0.401 & 36.6$\pm7.3$ &    0.90($+0.12,-0.17$) & Y \\
F36521\_1457 &   0.358 & 46.9$\pm4.7$ &    0.60($+0.10,-0.15$) \\
F36527\_1355 &   1.355 & 114.0($+28.5,-57$) & 2.25($+0.13,-0.36$) \\
F36534\_1140 &   1.275 & 54.7($13.7,-27.4$) &  1.11($+0.13,-0.34$) & Y \\
F36573\_1026 &   0.847 & 20.1($+5.0,-10.1$) &  0.29($+0.13,-0.36$) \\
F36599\_1450 &   0.762 & 35.0$\pm5.3$ &    1.47($+0.11,-0.15$) & Y \\
F37004\_1617\tablenotemark{c}
      &   0.913 & 34.2$\pm5.1$ &   0.91($+0.11,-0.15$) \\
F37144\_1221 &   1.084 & 22.8($+5.7,-11.4$) &   0.65($+0.13,-0.36$) \\
$\cal{I}$ Galaxies \\
F36230\_1346 &   0.485 & 23.4$\pm2.3$ & 0.88$+0.10,-0.14$  \\
F36345\_1213 &   1.015 & 14.1$\pm1.4$ & 1.32$+0.10,-0.14$ & Y \\
F36397\_1010 &   0.509 & 14.0$\pm2.0$ & 0.23$\pm0.12$ \\
F36399\_1250\tablenotemark{c} 
    &   0.848 & 17.2$\pm1.7$ & 0.93$+0.10,-0.14$  \\
F36422\_1545 &   0.857 & 13.0$\pm2.8$ & 1.18$+0.12,-0.17$ &  Y \\
F36431\_1109\tablenotemark{c} 
    &   0.297 & 13.5($+6.3,-3.4$) & $-0.75\pm0.20$  \\
F36481\_1309 &   0.476 & 21.8$\pm2.2$ &  0.86($+0.11,-0.15$)   \\
F36509\_1031 &   0.410 & 10.4$\pm1.6$ &  0.33($0.11,-0.15$)  & Y \\
F36588\_1638 &   0.299 & 11.3$\pm1.7$ &  0.57($+0.11,-0.15$)  \\
F36588\_1435 &   0.678 & 4.7$\pm1.4$ &   0.49($0.14,-0.21$)  \\
F37020\_1123 &   0.136 & 22.1$\pm3.3$ &  0.20($+0.10,-0.14$)   \\
F37027\_1543 &   0.514 & 8.3$\pm1.2$ &  0.71($+0.11,-0.15$)   \\
F37046\_1429 &   0.561 & 8.8$\pm1.8$ &   0.52($+0.12,-0.17$) \\
F37058\_1154 &   0.904 & 23.3$\pm4.7$ &  1.26($+0.11,-0.17$)  & Y \\
F37083\_1056 &   0.423 & 34.5$\pm6.9$ &  1.39($+0.12,-0.17$)  & Y \\
F37138\_1424 &   0.475 & 34.2$\pm5.1$ &  0.79($+0.11,-0.15$)   \\
\enddata
\tablenotetext{a}{Rest frame equivalent width of 3727~\AA\ emission line.}
\tablenotetext{b}{These galaxies have been detected with the VLA
\citep{richards98,richards00}.}
\tablenotetext{c}{No axis ratio could be measured for this galaxy.}
\end{deluxetable}

\clearpage

\begin{deluxetable}{lcccc rcc}
\tablenum{4}
\tablewidth{0pt}
\small
\tablecaption{Medians of Samples \label{table_median}}
\tablehead{ 
\colhead{Sample} & \colhead{No.} & $z$ &
\colhead{SFR$^i$(H$\alpha$)} &
\colhead{SFR$^i$(3727)}
& \colhead{log[$M$]\tablenotemark{a}} & 
\colhead{SFR$^i$/10$^{11}M$\subsun} &
\colhead{SFR$^i$/10$^{11}M$\subsun} \\ 
\colhead{}  & \colhead{Gals.} & \colhead{}  & \colhead{($M$\subsun/yr)} &
\colhead{($M$\subsun/yr)} &
\colhead{($M$\subsun)}    &  \colhead{($M$\subsun/yr)(H$\alpha$)} 
& \colhead{($M$\subsun/yr)(3727)}  }
\startdata
HDF Sample \\
Xray+Radio+CFGRS\tablenotemark{b} & 11 & 0.76 
      & \nodata &  15.1 &   11.09 & \nodata & 13.8  \\
Xray+CFGRS/$\cal{E}$ & 22 & 0.84 & \nodata & 7.6 &  10.71 & \nodata & 17.8  \\ 
Xray+CFGRS/$\cal{I}$ & 16 & 0.49 & \nodata & 5.6 & 10.88 & \nodata & 5.4  \\ 
Local Galaxies \\
Local calib. & 14 & 0.00 & 2.9 &  \nodata & 10.75 & 7.2 & \nodata \\
\enddata
\tablenotetext{a}{Inferred from the luminosity at rest frame $K$.}
\tablenotetext{b}{Only galaxies with optical emission lines are included.}
\end{deluxetable}

\begin{deluxetable}{lccc rcc}
\tablenum{5}
\tablewidth{0pt}
\small
\tablecaption{Second Highest Value of Samples \label{table_extreme}}
\tablehead{ 
\colhead{Sample} & \colhead{No.Gals} & 
\colhead{SFR(H$\alpha$)} &
\colhead{SFR(3727)}
& \colhead{log[$M$]\tablenotemark{a}} & 
\colhead{SFR/10$^{11}M$\subsun} &
\colhead{SFR/10$^{11}M$\subsun} \\ 
\colhead{}  & \colhead{}  & \colhead{($M$\subsun/yr)} &
\colhead{($M$\subsun/yr)} &
\colhead{($M$\subsun)}    &  \colhead{($M$\subsun/yr)(H$\alpha$)} 
& \colhead{($M$\subsun/yr)(3727)}  }
\startdata
HDF Sample \\
Xray+Radio+CFGRS\tablenotemark{b} & 11 
    & \nodata &  29.4 & 11.55 & \nodata & 36.7 \\
Xray+CFGRS/$\cal{E}$ & 22 &  \nodata & 39.9 &  11.41 & \nodata & 83.2  \\
Xray+CFGRS/$\cal{I}$ & 16 & \nodata & 29.4 & 11.55 & \nodata & 14.3  \\
Local Galaxies \\
Local calib. & 14 &  53.7 &  \nodata & 11.15 & 40.5 & \nodata \\
\enddata
\tablenotetext{a}{Inferred from the luminosity at rest frame $K$.}
\tablenotetext{b}{Only galaxies with optical emission lines are included.}
\end{deluxetable}

\clearpage

\clearpage

\clearpage

\clearpage

\clearpage

\clearpage

\clearpage

\clearpage

\clearpage


\begin{thebibliography}{}
\bibitem[Alexander \etal(2003)]{alexander03}
Alexander, D.~M. \etal, 2003, \apj\ (in press) 
(Astro-ph/0304392)

\bibitem[Aussel \etal(1999)]{aussel99}
Aussel, H., Cesarsky, C.J., Elbaz, D. \& Starck, J.L., \aap, 342, 313

\bibitem[Baldwin, Phillips \& Terlevich(1981)]{baldwin81}
Baldwin, J.~A., Phillips, M.~M. \& Terlevich, R., 1981, \pasp, 93, 5

\bibitem[Barger \etal(2002)]{barger02} Barger, A.~J., Cowie, L.~L.,
Brandt, W.~N., Capak, P.,
Garmire, G.~P. Hornschemeier, A.~E., Steffen, A.~T. \&
Wehner, E.~H., 2002, \aj, 124, 1839

\bibitem[Barger \etal(2003)]{barger03} Barger, A.~J., Cowie, L.~L.,
Capak, P., Alexander, D.~M., Bauer, F.~E., Brandt, W.~N.,
Garmire, G.~P. \& Hornschemeier, A.~E.,
2003, \apj, 584, L61

\bibitem[Bell \& Kennicutt(2001)]{bell01}
Bell, E.~C. \& Kennicutt Jr., R., 2001, \apj, 548, 681

\bibitem[Bell(2002)]{bell02} Bell, E.~F., 2002, \apj, 577, 150

\bibitem[Bell(2003)]{bell03} Bell, E.~F., 2003, \apj, 586, 794

\bibitem[Buat \etal(2002)]{buat02}
Buat, V., Boselli,  A., Gavazzi,G. \& Bonfanti, C., 2002, 
\aap, 383, 801


\bibitem[Calzetti \etal(1994)]{calzetti94} Calzetti, D.
1994, \apj, 429, 582

\bibitem[Calzetti \etal(1995)]{calzetti95} Calzetti, D.
1995, \apj, 443, 136

\bibitem[Calzetti(1999)]{calzetti99} Calzetti, D., 1999,
Astrophysics and Space Science, 266, 243

\bibitem[Cardiel \etal(2003)]{cardiel03}
Cardiel, N., Elbaz, D., Schiavon, R.~P., Willmer, C.~N.~A., 
Koo, D.~C., Phillips, A.~C. \& Gallege, J., 2003,
\apj, 584, 76

\bibitem[Cohen \etal(1996)]{cohen96} 
Cohen, J.~G., Cowie, L.~L., Hogg, D.~W., Songaila, A.,
Blandford, R., Hu, E.~M., \& Shopbell, P.,
1996, \apj, 471, L5 (C96)


\bibitem[Cohen \etal(1999)]{cohen99}
Cohen, J.~G., Hogg, D.~W., Pahre, M.~A., Blandford, R., 
Shopbell, P.~L. \& Richberg, K., 1999, \apjs, 120, 171


\bibitem[Cohen \etal(2000)]{cohen00} 
Cohen, J.~G.,  Hogg, D.~W., Songaila, A.,
Blandford, R., Cowie, L.~L., Hu, E.~M., Shopbell, P.
\& Richberg, K., 2000, \apj, 538, 29

\bibitem[Cohen(2001)]{cohen01} Cohen, J.~G., 2001, \aj,
121, 2895

\bibitem[Cohen(2002)]{cohen02} Cohen, J.~G., 2002,
\apj, 567, 672

\bibitem[Cohen(2003)]{cohen03} Cohen, J.~G., 2003,
to be published in {\it{Galaxy Evolution: Theory and
Observations}}, ed. V. Avila-Reese, C.Firmani, C.Frenk
\& C. Allen, RevMexAASC

\bibitem[Condon(1992)]{condon92} Condon, J.~J., 1992,
\araa, 30, 575

\bibitem[Cowie \etal(1996)]{cowie96}
Cowie, L.~L., Songaila, A., Hu, E.~M. \& Cohen, J.~G., 1996, 
\aj, 112, 839

\bibitem[Fabbiano(1989)]{fabbiano89}
Fabbiano, G., 1989, \araa, 27, 139

\bibitem[Fabbiano, Kim \& Trinchieri(1992)]{fabbiano92}
Fabbiano, G., Kim, D.-W. \& Trinchieri, G., 1992, \apjs, 80, 531

\bibitem[Fabbiano, Zezas \& Murray(2001)]{fabbiano01}
Fabbiano, G., Zezas, A. \& Murray, S.~S., 2001, \apj, 554, 1035

\bibitem[Ferrarese \& Merritt(2000)]{ferrarese00}
Ferrarese, L. \& Merritt, D., 2000, \apj, 539, L9


\bibitem[Gallagher, Bushouse \& Hunter(1989)]{gallagher89}
Gallagher, J.~S., Bushouse, H. \& Hunder, D.~A.,
1989, \aj, 97 700

\bibitem[Garrett(2002)]{garrett02}  Garrett, M.~A., 2002,
\aap, 384, L19

\bibitem[Giavalisco \etal(2003)]{giavalisco03}
Giavalisco, M. \etal, 2003, \apj, submitted

\bibitem[Grimm, Gilfanov \& Sunyaev(2003)]{grimm03}
Grimm, H.~J., Gilfanov, M. \& Sunyaev, R., 2003, \mnras, 339, 793

\bibitem[Ho, Filippenko \& Sargent(1997)]{ho97}
Ho,  L., Filippenko, A.~V. \& Sargent, W.~L.~W., 1997,
\apj, 487, 568

\bibitem[Hogg \etal(1998)]{hogg98}
Hogg D.~W., Cohen J.~G., Blandford R. \& Pahre, M.~A., 1998,
\apj, 504, 622

\bibitem[Hogg \etal(2000)]{hogg00} 
Hogg D.~W., Pahre M.~A., Adelberger K.~L., Blandford R., Cohen J.~G.,
Gautier T.~N., Jarrett T., Neugebauer G. \& Steidel C.~C., 2000, 
\apjs, 127, 1

\bibitem[Hornscheimeier \etal(2003)]{horn03}
Hornscheimier, A.~E. \etal, 2003, \apj\ (in press), see also Astro-ph/0305086

\bibitem[Huchra \& Burg(1992)]{huchra92} Huchra, J.~P., \& Burg, R.,
1992, \apj, 393, 90

\bibitem[Immler, Pietsch \& Aschenback(1998)]{immler98}
Immler, S., Pietsch, W. \& Aschenback, B., 1998, \aap, 331, 601

\bibitem[Jansen, Franx \& Fabricant(2001)]{jansen01}
Jansen, R.~A., Franx, M. \& Fabricant, D., 2001, \apj, 551, 825

\bibitem[Jarrett \etal(2003)]{jarrett03}
Jarrett, T.~H., Chester, T., Cutri, R., Schneider, S. \& Huchra, J.~P.,
2003, \aj, 125, 525

\bibitem[Kauffmann \etal(2003)]{kauffmann03}
Kauffmann, G. \etal, 2003, \mnras, 341, 54

\bibitem[Kennicutt(1983)]{kennicutt83} Kennicutt Jr., R.~C., 1983,
\apj, 272, 54

\bibitem[Kennicutt(1992)]{kennicutt92} Kennicutt Jr., R.~C., 1992,
\apj, 388, 310

\bibitem[Kennicutt(1998)]{kennicutt98} Kennicutt Jr., R.~C., 1998,
\araa, 36, 189

\bibitem[Kobulnicky \etal(2003)]{kobulnicky03} 
Kobulnicky, H.~A. \etal, 2003, \apj, submitted (Astro-ph/0305024)

\bibitem[Lilly \etal(1996)]{lilly96}
Lilly, S.~J., LeFevre, O., Hammer, F. \& Crampton, D.~C., 1996,
\apj, 460, L1

\bibitem[Lira \etal(2002)]{lira02} Lira, P., Ward, M.,
Zeza, A., Alonso-Herrero, A. \& Ueno, S., 2002, \mnras, 330 259

\bibitem[Madau, Pozzetti \& Dickinson(1998)]{madau98}
Madau, P., Pozzetti, L. \& Dickinson, M., 1998, \apj, 498, 106

\bibitem[Makishima \etal(2000)]{makishima00}
Makishima, K. \etal, 2000, \apj, 535, 632

\bibitem[Martin, Kobulnicky \& Heckman(2002)]{martin02}
Martin, C.~L., Kobulnicky, H.~A. \& Heckman, T.~M., 2002, \apj, 574, 663

\bibitem[McQuade, Calzetti \& Kinney(1995)]{mcquade95}
McQuade, K., Calzetti, D. \& Kinney, A.~L.,1995, \apjs, 97, 331

\bibitem[Moran, Lehnert \& Helfand(1999)]{moran99}
Moran, E.~C., Lehnert, M.~D. \& Helfand, D.~J., 1999,
\apj, 526, 649

\bibitem[Moran, Filippenko \& Chornock(2002)]{moran02}
Moran, E.~C., Filippenko, A.~V., \& Chornock, R. 2002, \apj, 579, L71

\bibitem[Okada, Mitsuda \& Dotani(1997)]{okada97}
Okada, K., Mitsuda, K. \& Dotani, T., 1997, PASJ, 49, 653

\bibitem[Oke \etal(1995)]{oke95}
Oke, J.~B., Cohen, J.~G., Carr, M., Cromer, J.,  Dingizian, A., Harris, F.~H.,
Labrecque, S., Lucinio, R.,  Schaal, W., Epps, H. \&  Miller, J., 1995, \pasp,
107, 307



\bibitem[Phillips \etal(1997)]{phillips97}
Phillips, A..~C., Guzm\'an, R., Gallego, J., Koo, D.~C., Lowenthal, 
J.~D., Vogt, N.~P., Faber, S.~M. and Illingworth, G.~D. 1997, 
\apj, 489, 543

\bibitem[Ptak \& Griffiths(1999)]{ptak99} Ptak, A.
\& Griffiths, R., 1999, \apj, 517, L85

\bibitem[Read \& Stevens(2002)]{read02} Read, A.~M. \& Stevens,
I.~R., 2002, \mnras, 335, L36

\bibitem[Rola, Terlevich \& Terlevich(1997)]{rola97}
Rola, C.~S., Terlevich, E. \& Terlevich, R.~J., 1997, \mnras, 289, 419

\bibitem[Richards \etal(1998)]{richards98}
Richards, E.A., Kellermann, K.I., Fomalont, E.B., 
Windhorst, R.A. \& Partridge, R.B.,
1998, \aj, 116, 1039

\bibitem[Richards(2000)]{richards00}
Richards, E.~A., 2000, \apj, 533, 611

\bibitem[Rosa-Gonzalez, Terlevich \& Terlevich(2002)]{rosa02}
Rosa-Gonzal\'alez, D., Terlevich, E. \& Terlevich, R., 2002,
\mnras, 332, 283

\bibitem[Schlegel, Finkbeiner \& Davis(1998)]{schlegel98}
Schlegel, D.~J., Finkbeiner, D.~P. \& Davis, M., 1998, 
\apj, 500, 525

\bibitem[Schmidt(1959)]{schmidt59} Schmidt, M., 1959, \apj, 129, 243

\bibitem[Skrutskie \etal(1997)]{skrutskie97}
Skrutskie, M.~F., Schneider, S.E., Stiening, R., Strom, S.E.,
Weinberg, M.D., Beichman, C., Chester, T. \etal, 1997, in {\it{The
Impact of Large Scale Near-IR Sky Surveys}}, ed. 
F.Garzon \etal\ (Dordrecht: Kluwer), p. 187

\bibitem[Storchi-Bergmann, Kinney \& Challis(1995)]{storchi95}
Storchi-Bergmann, T., Kinney, A.~L. \& Challis, P., 1995, \apjs, 98, 103

\bibitem[Terashima \& Wilson(2001)]{terashima01}
Terashima, Y. \& Wilson, A.~S., 2001, \apj, 560, 139

\bibitem[Tenorio-Tagle \etal(1999)]{tenorio99}
Tenorio-Tagle G., Silich, S.~A., Kunth, D., Terlevich, E.
\& Terlevich, R., 1999, \mnras, 309, 332

\bibitem[Terashima \& Wilson(2001)]{terashima03}
Terashima, Y. \& Wilson, A.~S., 2003, \apj, see Astro-ph/0305563

\bibitem[Tully \etal(1998)]{tully98}
Tully, R.~B., Pierce, M.~J., Huang, J.~S., Saunders, W.,
Verheijen, M.~A.~W. \& Witchalls, P.~L., 1998,
\aj, 115, 2264

\bibitem[Ueda \etal(2001)]{ueda01}
Ueda , Y., Ishisaki, T., Makishima, K. \& Ohashi,T., 2001, \apjs, 133, 1

\bibitem[Veilleux \& Osterbrock(1987)]{veilleux87}
Veilleux, S. \& Osterbrock, D.~E., 1987, \apjs, 63, 295

\bibitem[Vogler \& Pietsch(1997)]{vogler97}
Vogler \& Pietsch, 1997, \aap, 319, 459

\bibitem[Williams \etal(1996)]{williams96}
Williams, R.~E., \etal\ 1996, \aj, 112, 1335

\bibitem[Wilson \etal(2002)]{wilson02}
Wilson, G., Cowie, L.~L., Barger, A.~J. \& Burke, D.~J., 2002,
\aj, 124, 1258

\bibitem[Young \etal(1996)]{young96} 
Young, J.~S., Allen, L., Kenney, J.~D.~P., Lesser, A.
\& Rownd, B., 1996, \aj, 112, 1903

\bibitem[Zezas \& Fabbiano(2002)]{zezas02}
Zezas, A. \& Fabbiano, G., 2002, \apj, 577, 726

\end{thebibliography}
\end{document}